\documentclass[tighten, twocolumn]{aastex62}

\hypersetup{linkcolor=blue,citecolor=blue}

\newcommand{\nrao}{National Radio Astronomy Observatory, 520 Edgemont Rd, Charlottesville, VA, 22903, USA}
\newcommand{\haystack}{Massachusetts Institute of Technology, Haystack Observatory, 99 Millstone Rd, Westford, MA 01886, USA}
\newcommand{\bhi}{Black Hole Initiative, Harvard University, 20 Garden Street, Cambridge, MA 02138, USA}
\newcommand{\naoj}{Mizusawa VLBI Observatory, National Astronomical Observatory of Japan, 2-21-1 Osawa, Japan}
\newcommand{\kyoto}{Department of Astronomy, Kyoto University, Kitashirakawa, Oiwake-Cho, Sakyo-ku, Kyoto 606-8502, Japan}
\received{January 1, 2018}
\revised{January 7, 2018}
\accepted{\today}
\submitjournal{ApJ}

\begin{document}
\title{
Black hole Spin Measurement Based on Time-domain VLBI Observations of Infalling Gas Cloud}
\author{Kotaro Moriyama}
\affil{\naoj}
\affil{\haystack}
\email{moriyama@kusastro.kyoto-u.ac.jp}
\author{Shin Mineshige}
\affil{\kyoto}
\author{Mareki Honma}
\affil{\naoj}
\author{Kazunori Akiyama}
\affil{\naoj}
\affil{\haystack}
\affil{\nrao}
\affil{\bhi}

\begin{abstract}
The black hole spacetime is described by general relativity and characterized by two quantities: the black hole mass and spin.
Black hole spin measurement requires information from the vicinity of the event horizon, which is spatially resolved for the Galactic center Sagittarius\,A* (Sgr\,A*) and nearby radio galaxy M\,87 by means of very long baseline interferometry (VLBI) observations with the Event Horizon Telescope (EHT).
In this paper, we simulate EHT observations for a gas cloud intermittently falling onto a black hole, and construct a method for spin measurement based on its relativistic flux variation.
The light curve of the infalling gas cloud is composed of peaks formed by photons which directly reach a distant observer and by secondary ones reaching the observer after more than one rotation around the black hole.
The time interval between the peaks is determined by a period of photon rotation near the photon circular orbit which uniquely depends on the spin. 
We perform synthetic EHT observations for Sgr\,A* under a  more realistic situation that a number of gas clouds intermittently fall towards the black hole with various initial parameters.
Even for this case, the black hole spin dependence is detectable in correlated flux densities which are accurately calibrated by baselines between sites with redundant stations.
The synthetic observations indicate that our methodology can be applied to EHT observations of Sgr\,A* since April 2017.

\end{abstract}
\keywords{Accretion --- Black hole physics --- Gravitation --- Radiative transfer --- Relativity}

\section{Introduction}\label{sec_intro}

The black hole is one of the most eccentric objects predicted by general relativity.
The scale and geometry of the black hole spacetime is uniquely determined by its mass and spin.
The mass can be accurately estimated, for instance, by measuring orbits of stars inside the sphere of its gravitational influence [e.g., \citealp{Gravity2018} for the supermassive black hole Sagittarius\,A* (Sgr\,A*) at the Galactic center] or measuring the diameter of the photon ring encircling its black hole shadow (\citealp{EHT2019a, EHT2019f} for M87* at the nearby radio galaxy M87).
%
In contrast, the spin measurement is not easy, since it requires information 
around the marginally stable orbit (e.g., \citealp{Brenneman2006, Reynolds2014, Vasudevan2016}), and/or the vicinity of the black hole (e.g., \citealp{Abbott2017}).

Such close vicinity of a black hole is now accessible with the Event Horizon Telescope (EHT), a global very long baseline interferometry (VLBI) array observing at 1.3~mm (\citealp{Doeleman2009, EHT2019b}). 
EHT has resolved horizon-scale emission near the supermassive black holes Sgr A* (\citealp{Doeleman2008, Fish2011, Johnson2015, Fish2016, Lu2018}) and M87* (\citealp{Doeleman2012, Akiyama2015}).
Especially, the first-ever horizon-scale image of a black hole was provided by recent EHT observations of M87*, 
demonstrating the uniqueness of the EHT for probing the black hole spacetime by spatially and timely resolving the electromagnetic emission around the event horizon (\citealp{EHT2019a,EHT2019b,EHT2019c,EHT2019d,EHT2019e,EHT2019f}). 

Sgr\,A* has the largest angular size of the gravitational radius $1~r_{\rm g}\sim 5~{\rm \mu as}$ ($r_{\rm g}=GM/c^2$, where $G$ is the gravitational constant, and $c$ is the speed of light), given by its black hole mass of $M \sim 4\times10^6~M_{\odot}$ and Earth-Sgr\,A* distance of  $\sim 8$~kpc (e.g., \citealp{Ghez2008, Gillessen2009, Reid2004, Gravity2018}).
It is an extremely low luminous black hole candidate ($L\sim 10^{-9}~L_{\rm Edd}$, where $L_{\rm Edd}$ is the Eddington luminosity), and therefore has a low mass accretion rate of $\dot{M}\sim10^{-8}~M_{\odot}{\rm yr}^{-1}$ (\citealp{Agol2000, Bower2000, Marrone2007, Bower2018}). 
The continuum spectrum and the above properties of Sgr\,A* can be well explained by radiative inefficient accretion flow (RIAF) models (e.g., \citealp{Narayan1995, Narayan2000, Yuan2003, Yuan2009, Yuan2014, Maryam2018}).

A unique radiative property of Sgr A*, compared with that of M87*, is rapid time variabilities observed by a large number of multi-wavelength observations.
The timescale of the complex flux variation at radio, Near-Infrared, and X-ray frequency is 10 minutes through hours which is comparable to a rotation period at the marginally stable orbit (e.g., \citealp{Baganoff2001, Genzel2003, Dodds-Eden2009, Marrone2008, Yusef-Zadeh2009, Neilsen2013}).
With full-array observations of the EHT since 2017, horizon-scale imaging of dynamical structures of accretion flow near the event horizon of Sgr\,A* will be available (\citealp{Johnson2017, Bouman2017}), and many approaches of time-domain analysis have been actively developed (e.g., \citealp{Shiokawa2017, Medeiros2018}).

Black hole spin measurement of Sgr\,A* based on VLBI observations has been motivated by numerous theoretical studies of the spacetime dependence of the black hole shadow and photon ring (e.g., \citealp{Luminet1979, Fukue1988, Falcke2000, Takahashi2004, Bambi2009, Johannsen2016, Younsi2016}).
The vast majority of previous work infers the spin by modeling interferometric EHT data sets with snapshot images of numerical simulations (\citealp{Huang2009, Moscibrodzka2009, Dexter2010, Shcherbakov2012, Chan2015, Broderick2016, Chael2018b}), or semi-analytic models (\citealp{Huang2009, Broderick2011, Broderick2016}) together with other observational properties.
These methods are partly successful, though we wish to note that each approach has some uncertainties for electron heating prescriptions and selected initial conditions (e.g., \citealp{Pu2016, Pu2018, Dexter2010, Frage2016, Chael2018b}).

We should also point out that spin values deduced from above approaches have been inconsistent.
The spin values from semi-analytical models are $a/M\lesssim 0.9$ (\citealp{Huang2009}) and $a/M=0-0.4$ (\citealp{Broderick2016}), while those of the GRMHD simulations are $a/M \sim 0.9$ (\citealp{Moscibrodzka2009}) and $a/M\sim 0.5$ (\citealp{Shcherbakov2012}), where $a/M$ is the normalized spin parameter.
Therefore, an independent method which is less affected by accretion flow models is essential to constrain the spin of the black hole with EHT observations.

As an alternative approach, we have  investigated potential methodologies utilizing time variations in the brightness distribution of a gas cloud intermittently falling onto the black hole, while \citet{Broderick2005} investigate flux variation of a compact emitting region rotating around a black hole on a circular orbit. We have proposed new methods for measuring the black hole spin and deviation from the Kerr metric (\citealp{Moriyama2015, Moriyama2016}).
In these studies, we proposed them based on gradual flux increases due to the focusing effect around the photon circular orbit under an ideal situation that the infalling gas cloud has the critical angular momentum of the particle rotating on the marginally stable orbit.
We emphasize that this may be too tight assumption for practical applications; the gas cloud with a rotation velocity slower than Keplerian one may not generate such a gradual increase, since the gas cloud rapidly traverses the photon circular orbit and falls onto the black hole.
Furthermore, the assumption of the rotational velocity will be inconsistent with results of RIAF models and GRMHD simulation for Sgr\,A*, since they indicate accretion flow with sub-Keplerian orbital velocities (\citealp{Manmoto1997, Huang2009, Yuan2009, Narayan2012}).

In this study, we simulate EHT observations for infalling gas clouds around the black hole, and improve our methodology towards a more practical one. 
We  calculate a motion of an infalling gas cloud with various model parameters and photon trajectories, and investigate the spin dependence of relativistic flux variation using general relativistic ray-tracing.
We perform synthetic EHT observations for Sgr A* under a situation that a number of gas clouds intermittently fall onto the black hole, and propose a method for black hole spin measurement.
Our new methodology is based on the spatially-resolved temporal variation of infalling gas clouds to be observed with the EHT, and therefore may provide an independent spin estimate for Sgr\,A*.

The plan of this paper is as follows.
In Section \ref{sec_method}, we describe our methods 
of calculating observed images and flux variation from an infalling gas cloud. 
In Section \ref{sec_results}, we show the relativistic features of the spatial snapshot images and temporal flux variation of the infalling gas cloud, and then construct the method for black hole spin measurement.
In Section \ref{sec_discussion}, we discuss the observational feasibility based on EHT observations expected in 2017-2020 and general situation of infalling gas clouds.
Finally in Section \ref{sec_summary} we summarize the results obtained in this paper and discuss future issues.

\section{Model and methods of numerical calculations}\label{sec_method}
\begin{figure}
\plotone{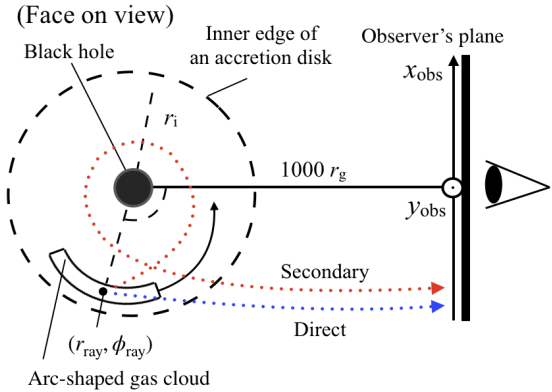}
\caption{
Schematic picture explaining the motion of an infalling gas cloud and observer's plane.
Here, $x_{\rm obs}$ and $y_{\rm obs}$ are Cartesian coordinates on the observer's plane, 
the $x_{\rm{obs}}$-axis is parallel to the equatorial plane of the black hole ($\otimes $), 
and the $y_{\rm{obs}}$-axis is perpendicular to the $x_{\rm{obs}}$-axis.
A ray emitted from a radial position, $r_{\rm ray}$, reaches the observer's plane with a rotational angle, $\phi_{\rm ray}$.
Further, we include in this figure the radiation directly reaching the observer (direct component, blue arrow) and that can reach the observer after more than one rotation around the black hole (secondary, red  arrow).
}\label{setting}
\end{figure}


\subsection{Simple gas cloud model}\label{subsec_simple}

We calculate relativistic flux variation emitted from gas clouds using our general relativistic ray-tracing scheme described in \citet{Moriyama2015}.
We consider a situation that gas clouds are tidally stripped off from the innermost part of the accretion disk. 
The gas clouds will get elongated due to radial differences in the azimuthal velocity and intermittently fall toward the black hole.
Then, it starts to fall onto a black hole with keeping its original angular momentum and hence the energy due to the relation of $u^{\mu} u_{\mu}=-1$, where $u^\mu$ is the four-velocity of the gas cloud (see Figure \ref{setting}).
We assume that the infalling gas cloud has magnetic field, and set the magnitude is $B\sim 100~{\rm G}$ from previous RIAF models or GRMHD simulations (\citealp{Yuan2009, Goldstone2005, Moscibrodzka2009, Dexter2010, Narayan2012, Chael2018b}).
The initial radius of the infalling gas cloud is set to be $0.98\, r_{\rm i}$, where $r_{\rm i}$ is the radius of the inner edge of the accretion disk.

To investigate relativistic flux variation, we construct a following simple gas cloud model which emits radiation with the power law spectrum. 
The emissivity in the inertial frame of the gas cloud is expressed by 
\begin{eqnarray}\nonumber
j (\nu_{\rm em}) &=& j_0 \exp\left[-\left(R/R_{\rm 0}(\phi) \right)^2\right]\nu_{\rm em}^{\Gamma+1}, \\ \label{eq_emissivity}
R_{0}(\phi) &=& R_{\rm cloud}\left[1-(\phi- \phi_{\rm cloud})/\delta\phi\right],
\end{eqnarray}
where $R$ is the radial distance from the central axis of the gas cloud, $R_{\rm cloud}$ is a characteristic thickness of the cross section, $\nu_{\rm em}$ is the photon frequency, $j_0$ is a numerical constant, $\Gamma$ is the photon index of the power-law spectrum.
The region of $[\phi_{\rm cloud}-\delta\phi$, $\phi_{\rm cloud}+\delta\phi]$ corresponds to the azimuthal region of the gas cloud, and $\phi$ is the azimuthal angle of the emission point. 
In the case of $\delta \phi \sim 1~ (\sim 2\pi)$, the gas cloud has arc (ring) shape.
The motion of the center of the gas cloud follows the one particle orbit given by geodesic equations (\citealp{Bardeen1972, Shapiro1983, Moriyama2015}). 
It falls to the black hole on the equatorial plane, keeping its original energy and angular momentum.
We summarize fiducial model parameters in Table \ref{table_parameters}.

\subsection{Gas cloud model based on synchrotron emission}\label{subsec_synch}
Many studies infer that the primary radio emission at the frequency of EHT observations ($\nu_{\rm  EHT}=230~{\rm GHz}$) around Sgr\,A* is the synchrotron radiation (\citealp{Narayan1995, Manmoto1997, Yuan2003, Yuan2009, Yuan2014}).
We construct an optically thin synchrotron emissivity model of the infalling gas cloud based on a relativistic Maxwellian distribution of electrons (\citealp{Pacholczyk1977, Narayan1995}): 
\begin{eqnarray}\nonumber
j (\nu_{\rm em}) &=& j_{\rm synch}(\nu_{\rm em}) \exp\left[-\left(R/R_{\rm 0} \right)^2\right], \\\label{eq_synchrotron} 
R_{0}(\phi) &=& R_{\rm cloud}\left[1-(\phi- \phi_{\rm cloud})/\delta\phi\right],
\end{eqnarray}
\noindent
with
\begin{eqnarray}\nonumber 
&& j_{\rm synch} = 4.43 \times 10^{-30} \frac{n_{\rm e}\nu_{\rm em}}{K_{2}(1/\theta_{\rm e})}I'(x_{\rm M})\ \ {\rm erg\ cm^{-3}s^{-1}\ str^{-1} Hz^{-1}},\\ \nonumber
&& x_{\rm M} = \frac{2\nu_{\rm em}}{3\nu_{0}\theta_{\rm e}{}^2}, \ \nu_{0} = \frac{eB}{2\pi m_{\rm e}c},\ \theta_{\rm e}= \frac{kT_{\rm e}}{m_{\rm e}c^2},\\ \label{eq_synchrotron_emissivity}
&& I'(x_{\rm M}) = \frac{4.0505}{x_{\rm M}^{1/6}}\left(1+\frac{0.40}{x_{\rm M}^{1/4}}+\frac{0.5316}{x_{\rm M}^{1/2}} \right)\exp(-1.8899x_{\rm M}^{1/3}), 
\end{eqnarray}
where $K_{2}$ is the modified Bessel function of the second kind, $k$ is the Boltzmann's constant, $B$ is the magnetic field, and $n_{\rm e}$, $e$, $m_{\rm e}$, and $T_{\rm e}$ are the number density, charge, mass, and temperature of the electron, respectively.
The past literature for RIAF models or GRMHD simulations show typical values of parameters as $\rho_{\rm gas}\sim 10^{-16}~{\rm g\ cm^{-3}}$, $T_{\rm e}\sim 10^{10}~{\rm K}$, and $B\sim 100~{\rm G}$, respectively (\citealp{Yuan2009, Goldstone2005, Moscibrodzka2009, Dexter2010, Narayan2012, Chael2018b}).

\subsection{Methods of numerical calculation}

Using the emissivity profile [Equations (\ref{eq_emissivity}) and (\ref{eq_synchrotron})], the radiative transfer equation is written as 
\begin{eqnarray}\label{radiation_transfer}
\Delta I(\nu_{\rm obs}) &=& g^{3}\int^{\nu_{\rm obs}+\Delta\nu}_{\nu_{\rm obs}}j(\nu/g) \Delta \ell d \nu  ,
\end{eqnarray}
where $I(\nu_{\rm obs})$ is the intensity of rays reaching the observer, $\nu_{\rm obs}$ is the observed photon frequency, $\Delta\nu$ is the width of the energy band for observations, $\Delta \ell$ is the infinitesimal spatial interval of a ray in Boyer-Lindquist coordinates.
For the 2017-2020 EHT observations, $\nu =230~ {\rm GHz}$, and $\Delta \nu~ \approx 2~{\rm GHz}$, respectively.
Further the energy-shift factor, $g$, is written as $g=p_0/(u_{\rm cloud}^{\mu}p_{\mu})$, where $u_{\rm cloud}^{\mu}$ is the four velocity of the emission point in the gas cloud, and $p_\mu$ is the four momentum of the emitted photon.
We fix the values of $\nu_{\rm obs}$ and $\Delta\nu$ as fundamental quantities of VLBI observations.

The actual calculation procedures are as follows:
\begin{enumerate}
\item Using the same procedures in \citet{Moriyama2017}, we numerically calculate the total intensity of rays, $I (x_{\rm obs},y_{\rm obs},t)$, that reach a cell at the Cartesian coordinates of $(x_{\rm obs},y_{\rm obs})$ on the observer's plane at the observer's time of $t$ (see Figure \ref{setting}).
Here, we take into account the slow-light effect to trace relativistic radiation transfer.
\item We integrate $I (x_{\rm obs},y_{\rm obs},t)$ over the observer's plane with the area and obtain the flux at each observer's time, $t$,
\begin{equation}\label{flux_equation}
f(t) =\frac{1}{4\pi D^2}\int I (x_{\rm obs},y_{\rm obs},t)dS_{\rm obs},
\end{equation}
where $dS_{\rm obs}$ is the area of the cell, and $D(= 1000r_{\rm g})$ denotes the distance between the center of the observer's plane and the black hole, and we focus on the radiation comes from the region of $r\leq r_{\rm i}+0.4r_{\rm g}$.
We have confirmed that numerical results remain the same, even if we take a longer distance of $D/r_{\rm g} = 10^4$. 
We use the normalized flux, $F(t)\equiv f(t)/f(t_{\rm max})$, where $t_{\rm max}$ denotes the time when $f(t)$ reaches its maximum.
\end{enumerate}

We divide the flux, $f(t)$, into two components. The first one directly reaches the observer (direct component, see blue arrow in Figure \ref{setting}), $f_0(t)$.
The second one is composed of photons that can reach the observer after more than one rotation around the black hole (secondary component, see red arrow in Figure \ref{setting}), $f_1(t)$.
To investigate the physical condition around the main emissivity region each time, $t$, we define the following average of a physical quantity, $P$: 
\begin{eqnarray}
\langle P(t)\rangle_{i} = \frac{1}{f_{i}(t)}\int P(x_{\rm obs},y_{\rm obs},t) df_{i}(x_{\rm obs}, y_{\rm obs}, t), 
\label{eq_average}
\end{eqnarray}
where $df=I(x_{\rm obs},y_{\rm obs},t)dS_{\rm obs}/(4\pi D^2)$, and the subscript, $i$($=$0 or 1) indicates the direct, or secondary component. 
Further, the value with no subscript corresponds to the averaged value of the total flux.

\section{Results}\label{sec_results}
In this section, we investigate the light fluctuation of the infalling gas cloud with the sub-Keplerian velocity using the simple gas cloud model (see Section \ref{subsec_simple}), and examine the relation between the black hole spin value and relativistic flux variation.

\begin{deluxetable*}{ccc}
\tablenum{1}
\tablecaption{Physical meanings and fiducial values of model parameters\label{table_parameters}}
\tablewidth{0pt}
\tablehead{
\colhead{Parameter} & \colhead{Physical meaning} & \colhead{Fiducial value for the gas cloud} 
}
\startdata
$R_{\rm cloud}/r_{\rm g}$ & Thickness of the cross section of the gas cloud & $0.2$\\ 
$\Gamma$ & Power-law index of the radiation spectrum & $-2$\\
$\delta\phi/\pi$ & Arc chord angle of the gas cloud & $2/3$\\
$i$ & Inclination angle & $75^\circ$ \\
$u_{\rm o}{}^r$ & Initial four-velocity of the radial component & $0$ \\ 
$L/L_{\rm ms}$ & Angular momentum of the gas cloud &  $0.8$\\
\enddata
\tablecomments{We denote $L_{\rm ms}$ as the angular momentum of the particle rotating on the marginally
stable orbit.}
\end{deluxetable*}


\begin{figure}\begin{center}\includegraphics[width=9cm]{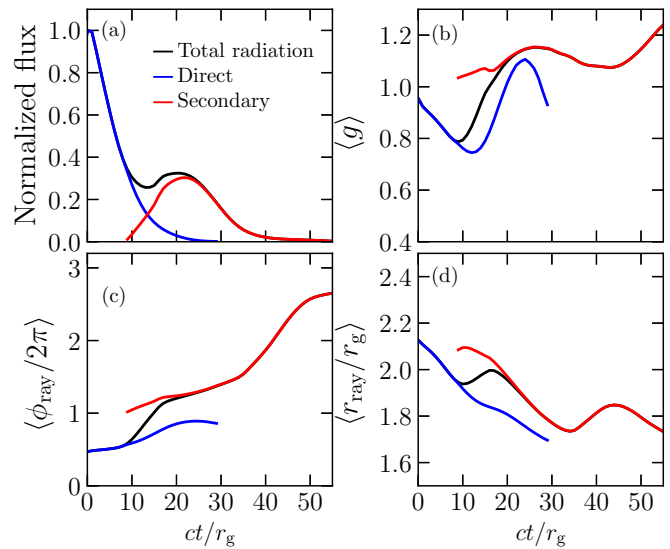} \end{center}
\caption{
Typical results of the infalling gas ring model with fiducial model parameters, $R_{\rm cloud}$, $\Gamma$, $\delta\phi$, $i$, $u_{\rm o}{}^r$, and $L$ that are summarized in Table \ref{table_parameters}, 
where we assign $(a/M,r_{\rm i})=(0.9,r_{\rm ms})$, $a/M$ is the normalized spin parameter, $r_{\rm i}$ is the radius of the inner edge of the accretion disk, and $r_{\rm ms}$ is the radius of the marginally stable orbit: 
(a) time variation of the normalized flux, $F$, 
(b) energy-shift factor, $\langle g\rangle$, 
(c) averaged radial position of the emission region, $\langle r_{\rm ray}/r_{\rm g}\rangle$, and
(d) averaged azimuthal phase, $\langle \phi_{\rm ray}/2\pi\rangle$ [the definition of averaged values is given by Equation (\ref{eq_average})].
Each panel shows values for total radiation (black curve), and those for direct (blue) and secondary (red) components.
}\label{ring}
\end{figure}

\subsection{Radiation properties of gas ring model}\label{subsec_ringmodel}

To understand the whole time variation of the radiation property of an infalling gas cloud, in Figure \ref{ring}, we show the time variation of the normalized flux and averaged values of the infalling gas ring model.
We assign fiducial model parameters summarized in Table \ref{table_parameters}, and $a/M=0.9, r_{\rm i}=r_{\rm ms}$, where $r_{\rm ms}$ is the radius of the marginally stable orbit. 
Each panel shows values for the total flux (black curve), and those for direct (blue) and secondary (red) components.

We show in Figure \ref{ring}a the light curve of the gas ring model.
The gas ring rapidly falls towards the event horizon before one rotation around the black hole, where the time duration for capturing the event horizon is $t_{\rm infall}=22 r_g/c$.
The total flux (the black curve) is dominated by the radiation of the direct component (blue curve) and secondary one (red curve), where we normalize each component by the maximum value of the total flux $\left[f(t=0)\right]$.
During $0\leq ct/r_{\rm g}<10$, the total flux is equal to that of the direct component, and rapidly decreases due to the gravitational redshift and to the capture of rays by the black hole.
After $ct/r_{\rm g}=10$, the light curve of the total flux has a transition from the direct component to the secondary one ($10\leq ct/r_{\rm g}<20$), and has a peak of the secondary component at $ct/r_{\rm g}=20$.
During $20\leq ct/r_{\rm g}<50$, the total flux is equal to that of the secondary component and slowly decreases due to the gravitational redshift.
In \citet{Moriyama2015}, we assumed that the gas ring has a high angular momentum ($L=L_{\rm ms}$) and showed that the flux first gradually increases due to the focusing effect around the photon circular orbit.
We emphasize that the gas cloud with the sub-Keplerian orbital velocity cannot generate such a gradual increase, since the gas cloud rapidly traverses the photon circular orbit and falls onto the black hole.

We investigate the averaged energy shift factor, $\langle g\rangle$, for each component (see Figure \ref{ring}b).
The energy shift factor of the direct component, $\langle g\rangle_{0}$ first decreases due to the gravitational redshift as the gas cloud approaches the black hole (see blue curve).
During $17\leq ct/r_{\rm g}\leq 24$, $\langle g\rangle_{0}$ increases, since the radiation from the high emissivity region is captured by the black hole, and that of the outer gas element, which has a low emissivity and modest $g$ due to the beaming effect, is dominant.
The energy shift factor of the secondary component, $\langle g\rangle_1$, is larger than that of the direct component and has gradual fluctuation within the range of $1-1.2$.
At $ct/r_{\rm g}=26$, $\langle g\rangle_{1}$ reaches its maximum after the time of the peak of the flux ($ct/r_{\rm g}=20$).

Next, we focus on the averaged azimuthal angle of the photon rotation $\langle \phi_{\rm ray}/2\pi\rangle$ [panel (c)], where the schematic definition of $\langle \phi_{\rm ray}/2\pi\rangle$ is shown in Figure \ref{setting}.
During $0\leq ct/r_{\rm g}\leq 10$, the azimuthal angle for the direct component, $\langle \phi/2\pi\rangle_0$, is within the range of $0.47-0.59$ due to the dominance of the combination of the beaming effect and gravitational lensing. 
As the gas cloud approaches the event horizon, $\langle \phi/2\pi\rangle_0$ increases due to the gravitational bending effect.
During $15\leq ct/r_{\rm g}\leq 30$, the azimuthal angle of the secondary component, $\langle \phi/2\pi\rangle_1=1.01-1.39$, and so the main radiation is the secondary component affected by the beaming effect.
At the time of $ct/r_{\rm g}>30$, $\langle \phi/2\pi\rangle_1$ rapidly increases, and the tertiary component ($\langle \phi/2\pi\rangle_1>2$) is dominant after the time of $ct/r_{\rm g}=42$.

Figure \ref{ring}d shows the time variation of the radial position of the main emission region of each component, $\langle r_{\rm ray}/r_{\rm g}\rangle_i$. 
The radial position for the direct component, $\langle r_{\rm ray}/r_{\rm g}\rangle_0$, monotonically decreases, since the gas ring falls to the black hole.
During $20\leq ct/r_{\rm g}\leq 33$, the secondary radiation is dominant, and $\langle r_{\rm ray}/r_{\rm g}\rangle_1$ also monotonically decreases due to the infalling gas ring.
After $ct/r_{\rm g}=30$, $\langle r_{\rm ray}\rangle_1$ increases and starts decreasing at $ct/r_{\rm g}=44$, since the secondary component decays and the tertiary one is dominant.
By comparing panels (a) and (d), we estimate the radius corresponding to the peak of the secondary component: $\langle r_{\rm ray}/r_{\rm g}\rangle_1|_{\rm max}=2.0$,  which is larger than that of the photon circular orbit, $r_{\rm ph}/r_{\rm g}=1.6$, since the emission at $r_{\rm ph}$ is small due to an increase in the fraction of photons that are captured by the black hole.

\begin{figure*}[t]
\begin{center}\includegraphics[width=17cm]{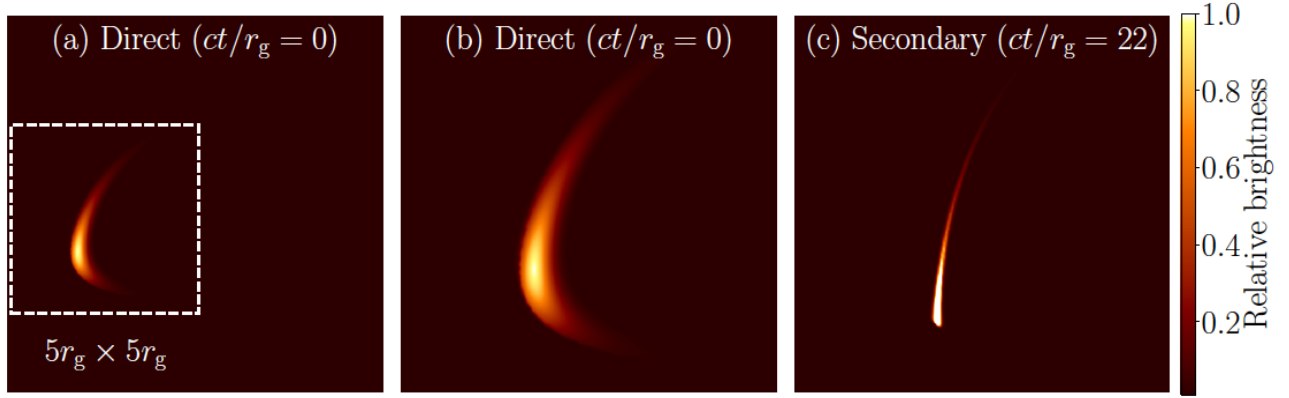} \end{center}
\caption{
Snapshots of the direct radiation [panels (a) and (b)] and secondary component [panel (c)] of a gas ring, where the parameters are the same as those in Figure \ref{ring}.
The abscissa is $x_{\rm obs}/r_{\rm g}$, while the ordinate is $y_{\rm obs}/r_{\rm g}$, and the color contours
represent the relative brightness of the gas-cloud image, $\Delta f(x_{\rm obs},
y_{\rm obs}, t)/\Delta f_{\rm norm}$, where the normalized brightness, $\Delta f_{\rm norm}$, is at the brightest position of the direct component, $(x_{\rm obs}/r_{\rm g}, y_{\rm obs}/r_{\rm g}) = (-3.1, 0.7)$. 
Here, the spatial regions of panels (b) and (c) correspond to the dotted rectangular frame region in panel (a).
}\label{ring_image}
\end{figure*}

In Figure \ref{ring_image}, we show snapshots of the direct and secondary components.
The abscissa and ordinate are the Cartesian coordinates on the observer's plane, $(x_{\rm obs},y_{\rm obs})$, and the color contours represent the relative brightness of the gas cloud image, $\Delta f(x_{\rm obs}, y_{\rm obs}, t)/\Delta f_{\rm norm}$, where the normalized brightness, $\Delta f_{\rm norm}$, is at the brightest position of the direct component located at $(x_{\rm obs}/r_{\rm g}, y_{\rm obs}/r_{\rm g}) = (-3.1, 0.7)$. 
Panels (a) and (b) show the relative brightness of the image of the direct component whose crescent shape arises from Doppler boosting/deboosting on the approaching/receding side of the gas ring, where the spatial region of panel (b) is indicated by dotted rectangular frame in panel (a).
The image of the secondary component has a narrow shape structure [panel (c), where the region is the same as that of panel (b)], and the brightest region, $(x_{\rm obs}, y_{\rm obs})=(-3.0, 0.4)$, is the similar to that of the direct one, since the apparent size of the gas cloud is expanded by the gravitational bending effect.

\subsection{Flux variation of arc-shaped gas clouds}\label{subsec_arcgas}
\begin{figure*}\begin{center}\includegraphics[width=18cm]{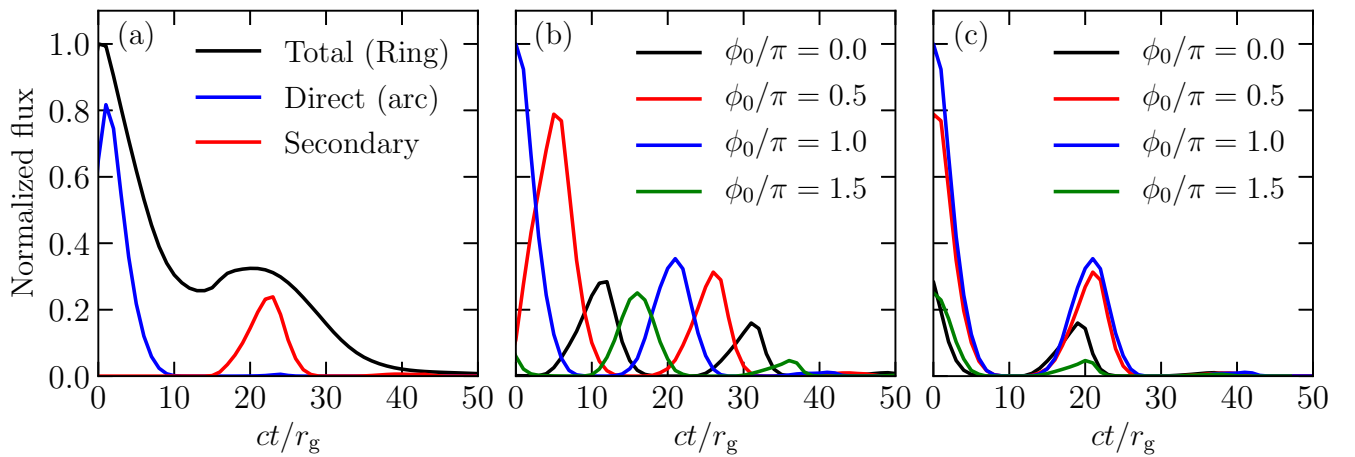} \end{center}
\caption{
Flux variation of the arc-shaped gas clouds, where the arc chord angle is $\delta \phi=2\pi/3$, $\phi_{\rm o}$ is the initial azimuthal angle, and other parameters are the same as those in Figure \ref{ring}.
Panel (a) shows the comparison of the light curve of the arc-shaped gas cloud with the initial azimuthal angle of, $\phi_{\rm o}/\pi =3/4$, (red and blue curves) and that of the gas ring (black), where the blue and red curves show flux variation of the direct and secondary components, respectively.
Panel (b) plots light curves of the arc-shaped gas cloud with the various initial azimuthal angle, 
$\phi_{\rm o}/\pi = 0,0.5,1.0,$ and $1.5$, where the flux is normalized by the maximum one in the case of $\phi_{\rm o}/\pi =1.0$.
Panel (c) depicts each light curve in panel (b) by aligning the maximum peaks.
}\label{arcgas}
\end{figure*}

In Figure \ref{arcgas}a, we show flux variation of an arc-shaped gas cloud with 
the initial azimuthal angle, $\phi_{\rm o}/\pi =3/4$ and fiducial values of model parameters listed in Table \ref{table_parameters}, where the blue (or red) curve indicates the radiation of the direct (secondary) component, and the black one corresponds to the total radiation of the ring model.
Here, if $\phi_{\rm o}=0$, the gas cloud position is in front of the black hole seeing from the observer.

At the time of $ct/r_{\rm g}=1$, the first peak appears due to Doppler boosting of the direct component.
The radial position of the peak is $\langle r_{\rm ray}/r_{\rm g}\rangle=2.1$ and the gas cloud reaches the event horizon after $22 r_{\rm g}/c$.
At the time of $ct/r_{\rm g}=23$, the second peak appears due to Doppler boosting of the secondary component, while the contribution of the direct component is negligible, since the gas cloud is captured by the black hole.
After the second peak, the third peak occurs at the time of $ct/r_{\rm g}= 43$, which is due to the tertiary component.
The radial positions of main radiation of second and third peak are $\langle r_{\rm ray}/r_{\rm g}\rangle=1.9$ and $1.8$, and the time intervals of each peak are $c\delta t/r_{\rm g}=22,$ and $20$ respectively.

In Figure \ref{arcgas}b, we show the light curves of the arc-shaped gas clouds with various initial azimuthal angles, $\phi_{\rm o}/\pi =0, 0.5, 1$, and $1.5$.
Each light curve has two or three peaks due to the direct and secondary components. 
We align the maximum peaks of the light curves and depict in Figure \ref{arcgas}c.
Each time interval between the direct and secondary radiation peaks is within the range of $19-21 r_{\rm g}/c$.

We emphasize that the time intervals between the two neighboring flux peaks are approximately a rotational period of photons near the photon circular orbit.
Using Equation (\ref{Omega_ray}) in the Appendix A, the angular velocity of the ray, $\Omega$, at the radial position $r/r_{\rm g}=1.9$ is $r_{\rm g}\Omega/c =0.33$, whose rotational period is $cT/r_{\rm g}=19$ from Equation (\ref{T_ray}).
We find that the time interval between each radiation component determines the lower limit of the flux variation for the dynamical motion of the gas cloud.
The timescale is consistent with the time intervals of the flux peaks of the infalling gas cloud.
In the next subsection, we focus on the fact that the radius of the photon circular orbit is uniquely determined by the spin value (\citealp{Bardeen1972}) and investigate the relation between the time interval and spin parameter.

\subsection{Black hole spin dependence of light curves of arc-shaped gas clouds}\label{subsec_spin_dependence}

\begin{figure*}\begin{center}\includegraphics[width=13cm]{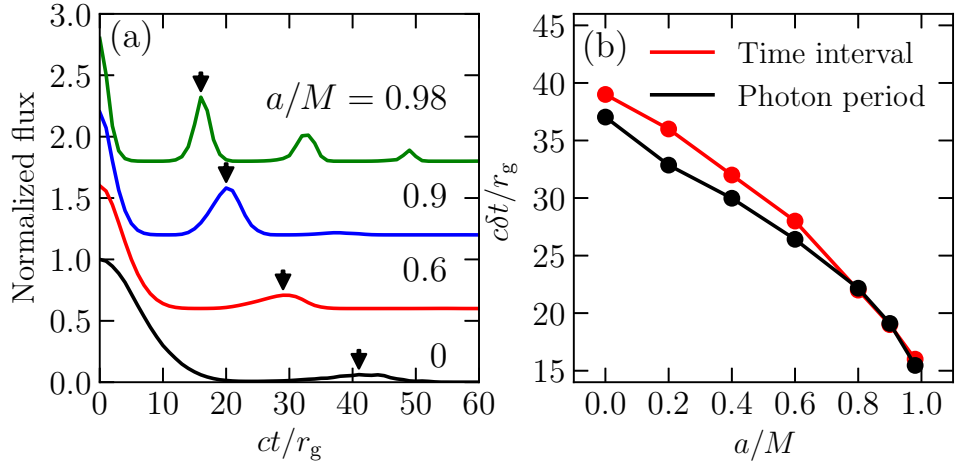} \end{center}
\caption{
Spin dependence of the superposed light curve of the arc-shaped gas cloud. 
We superpose light curve of the arc-shaped gas cloud with 20 initial azimuthal angles, $\phi_{\rm o}/(2\pi)=[0,19/20]$, by aligning their maximum peaks for the case of each spin [panel (a)].
The light curves from bottom to top are for $a/M=0$ (black curve), $0.6$ (red; shifted vertically by 0.6 for clarity), 
$0.9$ (blue; shifted vertically by 1.2), and $0.98$ (green; shifted vertically by 1.8), where the values of the other parameters are summarized in Table \ref{table_parameters}.
Panel (b) shows the spin dependence of the time interval between the first two peaks indicated by the red curve, while  the black curve shows the period for photon rotation, where the rotation radius is set to be the radial position of the main emission of the secondary component: $\langle r_{\rm ray}/r_{\rm g} \rangle =4.3, 3.7, 3.3, 2.9, 2.3, 1.9,$ and $1.4$ for $a/M=0, 0.2, 0.4, 0.6, 0.8, 0.9$, and $0.98$, respectively.
}\label{spin_dt}
\end{figure*}

As demonstrated in Section \ref{subsec_arcgas}, we assigned $a/M=0.9$ and $i=75^\circ$, and find that the time interval between the first two peaks due to the direct and secondary components, $\delta t$, is consistent with the rotational period of the rays around the photon circular orbit, $r_{\rm ph}$.

To investigate the spin dependence of $\delta t$, we superpose the light curve of the arc-shaped gas cloud with 20 initial azimuthal angles, $\phi_{\rm o}/(2\pi)=[0,19/20]$, by aligning their maximum peaks for the case of each spin (Figure \ref{spin_dt}a).
The light curves from bottom to top are for $a/M=0$ (black curve), $0.6$ (red; shifted vertically by 0.6 for clarity), 
$0.9$ (blue; shifted vertically by 1.2), and $0.98$ (green; shifted vertically by 1.8), where the other parameters are the same as those in Figure $\ref{arcgas}$.

Let us change the spin value from $a/M=0$ to $a/M=0.98$ to see the spin dependence of the superposed light curves.
The time width of the first peak of the direct component at $ct/r_{\rm g}\approx 0$ decreases with $a/M$.
This is because the gas cloud falls to the black hole from smaller radii (we note that the initial position at $r \approx r_{\rm ms}$ is smaller for a higher spin).
The higher $a$ is, the larger is the amplitude of the peak of the secondary component due to the frame-dragging effect.
The time interval between the first and second peaks decreases with spin, since the photon circular orbit decreases.
In the case of $a/M=0.98$, the light curve has four peaks due to the photon rotation around $r_{\rm ph}$. 
The radial positions of the main emissions of the secondary components are $\langle r_{\rm ray}/r_{\rm g}\rangle =4.3, 3.7, 3.3, 2.9, 2.3, 1.9,$ and $1.4$ for $a/M=0, 0.2, 0.4, 0.6, 0.8, 0.9$, and $0.98$, respectively.

We summarize in Figure \ref{spin_dt}b the spin dependence of time interval, $\delta t$ (red curve) defined by
\begin{eqnarray}\label{eq_dt}
\delta t = \left[\int^{t_{\rm f}}_{t_{\rm val}} tf(t) dt\right] \left[\int^{t_{\rm f}}_{t_{\rm val}}f(t) dt\right]^{-1} -t_{\rm dir},
\end{eqnarray}
where $t_{\rm dir}$ is the time at the peak of the direct component, $t_{\rm val}$ is the time of the valley between the direct and secondary components, and $t_{\rm f}$ is defined in such a way that $f(t)=f(t_{\rm val})$ $(t_{\rm f }>t_{\rm val})$.  
Further, the black curve shows that the period of photon rotation at the radius of the emission point of the secondary component. 
The time interval is uniquely determined by the black hole spin, and the period of photon rotation is consistent within the range of $3r_g/c$.

\begin{figure*}\begin{center}\includegraphics[width=13cm]{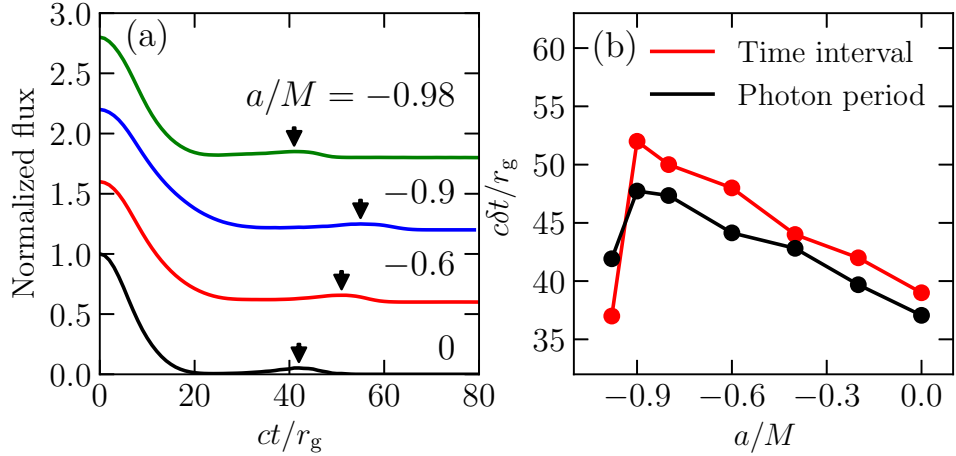} \end{center}
\caption{
The same as Figure \ref{spin_dt}, but for $a\leq 0$ case: 
the light curves from bottom to top are for $a/M=0$ (black curve), $-0.6$ (red; shifted vertically by 0.6 for clarity), 
$-0.9$ (blue; shifted vertically by 1.2), and $-0.98$ (green; shifted vertically by 1.8), where the values of the other parameters are summarized in Table \ref{table_parameters}.
In panel (b)  the black curve shows the period for photon rotation, where the rotation radius is set to be the radial position of the main emission of the secondary component: $\langle r_{\rm ray}/r_{\rm g}\rangle=
4.3, 4.6, 5.1, 5.1, 5.6, 5.6
$ and $5.7$ for $a/M=0, -0.2, -0.4, -0.6, -0.8, -0.9$, and $-0.98$, respectively.
}\label{spin_dt_ret}
\end{figure*}

Further, we evaluate the spin dependence of flux variation for the case with $a\leq 0$ (see Figure \ref{spin_dt_ret}a).
The light curves from bottom to top are for $a/M =$ 0 (black curve), -0.6 (red; shifted vertically by 0.6 for clarity), -0.9 (blue; shifted vertically by 1.2), and -0.98 (green; shifted vertically by 1.8), where the other parameters are the same as those in Figure \ref{arcgas}.
Except for extremely rotating black hole case, the gas cloud and photons for main radiation counter-rotate with the black hole spin, and the spin dependence of the features of direct and secondary components are explained from the consideration of co-rotating cases shown in Figure \ref{spin_dt}.
In the case of $a/M=0.98$, the photons co-rotating with the black hole spin are dominated due to a strong frame dragging effect.
The time interval between each peak is consistent with the photon rotation period at radial positions of the main emissions of the secondary components (see Figure \ref{spin_dt_ret}b).
For the counter-rotating case, the amplitude of the secondary component is quite small, and so hereafter we especially focus on the case of gas clouds co-rotating with the black hole.

\begin{figure*}\begin{center}\includegraphics[width=15cm]{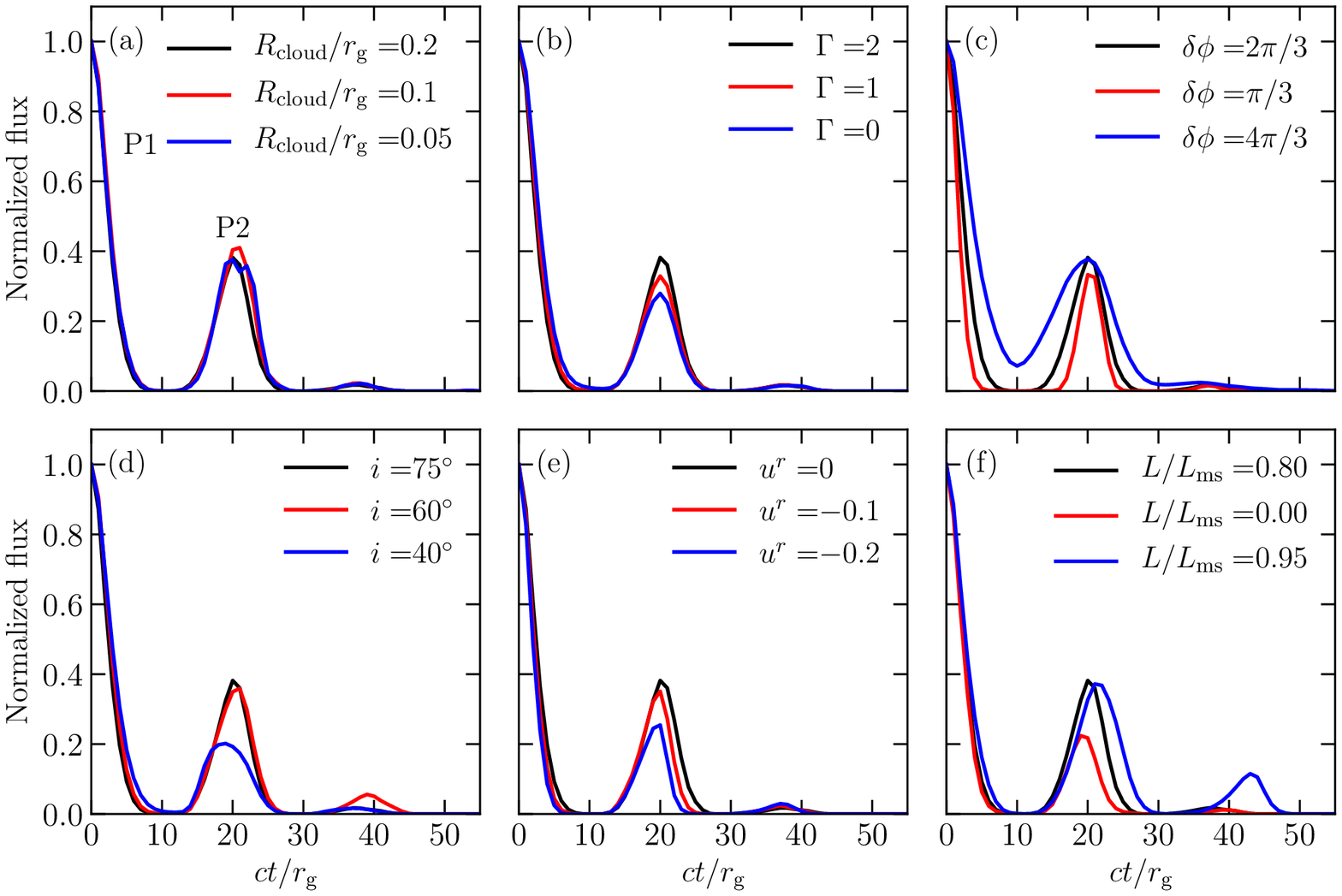} \end{center}
\caption{
Parameters dependence of the superposed light curve.
Panels (a)--(f) are the same as Figure \ref{spin_dt}a but for different values of $R_{\rm cloud}/r_{\rm g}=0.2, 0.1,$ and $0.05$, $\Gamma=-2, -1,$ and $0$, $\delta\phi=2\pi/3, \pi/3,$ and $4\pi/3$, 
$i=75, 60,$ and $40^\circ$, $u_{\rm o}{}^r=0, -0.1,$ and $-0.2$, and $L/L_{\rm ms}=0.8, 0,$ and $0.95$, respectively, where the physical meanings of each parameter are summarized in Table \ref{table_parameters}.
When not shown, the values of the parameters are fiducial ones summarized in Table \ref{table_parameters}.
We also denote $P_1$ and $P_2$ as the first and second peaks, respectively. 
}\label{parameter}
\end{figure*}
\subsection{Dependence of superposed light curves on model parameters}\label{subsec_superposed_light_curve}

In Section \ref{subsec_spin_dependence}, we focused on the radiative property of the arc-shaped gas cloud, and proposed a new method for the spin measurement using the time interval between the peaks of the direct and secondary components, $\delta t$.
One may think, however, that the results may sensitively depend on the model parameters: the shape of the gas cloud, emissivity profile, dynamical properties, and inclination angle.
To investigating the sensitivity, we need to consider how $\delta t$ depends on these parameters.
First, we focus on a high spin case ($a/M=0.9$), and show in Figure \ref{parameter} the dependences of model parameters: $R_{\rm cloud}, \Gamma, \delta\phi , i, u_{\rm o}{}^r,$ and $L/L_{\rm ms}$.
When not shown, the values of the parameters are fiducial ones listed in Table \ref{table_parameters}.
In this subsection, we denote $P_1$ and $P_2$ as the first and second peaks, respectively (see Figure \ref{parameter}a).

The time interval between the two peaks does not critically depend on the shape of the gas cloud, and emissivity profile (Figures \ref{parameter}a-c).
Figure \ref{parameter}a shows that the light curve does not depend on the thickness of the gas cloud, $R_{\rm cloud}$, and so we understand the direct and secondary components have similar dependence on $R_{\rm cloud}$.
The larger $\Gamma$ is, the weaker is the contribution of the energy-shift factor, $g$, to the intensity [see Equation (\ref{radiation_transfer})].
Since $g$ of the  secondary  component is larger than that of the direct one (see Figure \ref{ring}b), the amplitude of the second peak decreases with $\Gamma$ (Figure \ref{parameter}b).
We note that the range of $\Gamma$ in this study is derived from the result of the continuum spectrum of Sgr\,A* (see \citealp{Broderick2009}, and Equations (\ref{radiation_transfer}) and (\ref{flux_equation})].
If the arc chord angle, $\delta\phi$, increases, the peak intervals resemble each other, but the peak widths of $P_1$ and $P_2$ increase (Figure \ref{parameter}c). 
If we change the inclination angle from $i=75^\circ$ to $40^\circ$, the peak amplitude of $P_2$ decreases due to the decay of the photon-bending effect though the time interval between the two peaks not being significantly altered (Figure \ref{parameter}d).

Inside the radius of the marginally stable orbit, the RIAF model and GRMHD simulations, which well explains the continuum spectrum of Sgr\,A*, indicates the sub-Keplerian orbital velocity and 
moderate radial velocity ($\approx -0.1$, see \citealp{Manmoto1997, Yuan2009, Narayan2012}). 
We investigate the superposed light curves for the case of $u_{\rm o}{}^{r}=-0.1, -0.2$ and $L/L_{\rm ms}=0, 0.95$.
The larger $u_{\rm o}{}^r$, the time interval does not change significantly, but the amplitude of $P_2$ is smaller because of the redshift due to the radial motion of the gas cloud, (Figure \ref{parameter}e).
The peak amplitude of $P_2$ increases with $L/L_{\rm ms}$ due to the beaming effect by the high rotational velocity of the gas cloud (Figure \ref{parameter}f). 
By adapting the relation between $\delta t$ and $a/M$ to the superposed light curves in Figure \ref{parameter}, we find that the estimated spin is within the range of $= 0.83-0.93$, and the error is $|\delta a/M|<1.$.

\begin{figure*}\begin{center}\includegraphics[width=18cm]{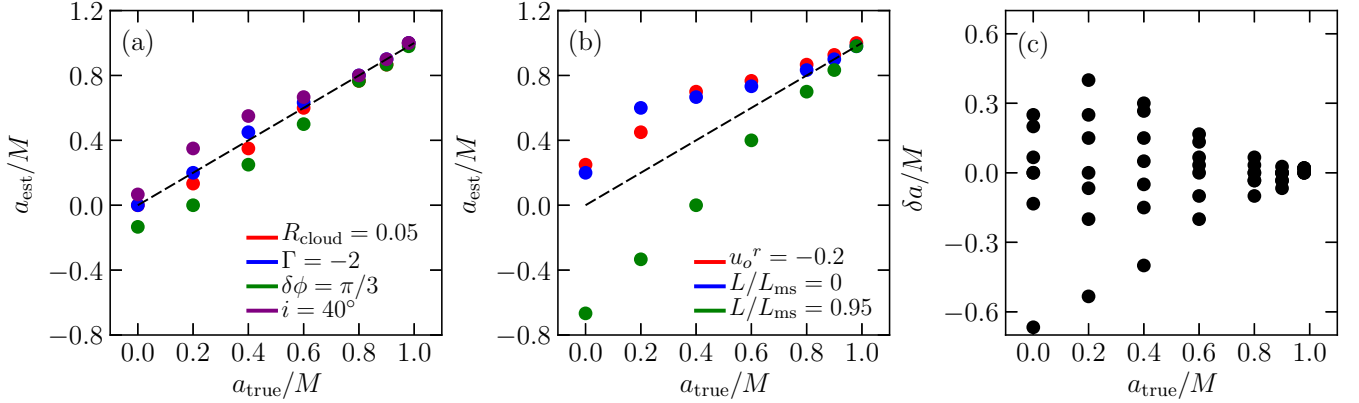} \end{center}
\caption{
Estimated range of the spin values by using the relation between the spin and time interval of the first two peaks, $c\delta t/r_{\rm g}$ shown in Figure \ref{spin_dt}b, where $a_{\rm true}$ and $a_{\rm est}$ are the actual and estimated spin values, respectively. 
Panel (a) shows the estimation of the spin values for the parameters of, $R_{\rm cloud}/r_{\rm g}=0.05$ (red cross), 
$\Gamma=2$ (blue square), and $\delta \phi=\pi/3$ (green circle), and $i=40^\circ$ (purple triangle).
Panel (b) plots the estimation of the spin values for the various values for $u_{\rm o}^{r}=0.2$ (red cross), 
$L/L_{\rm ms}=0$ (blue square), and $L/L_{\rm ms}=0.95$ (green circle), respectively.
For the case of $L/L_{\rm ms}=0$, we cannot estimate spins for $a_{\rm true}/M=0$ and $0.2$, since the second peak does not appear.
The fiducial parameters are shown in Table \ref{table_parameters}, and the dotted line represents the relation of $a_{\rm est}=a_{\rm true}$.
Panel (c) depicts the relation between real spin values and errors of the estimated spin, $\delta a/M = (a_{\rm est}-a_{\rm true})/M$ estimated in panels (a) and (b).
}\label{spin_spin}
\end{figure*}

Let us examine model dependencies of estimated spin values $a_{\rm est}/M$, by observing $\delta t$ for each value of true spin, $a_{\rm true}$.
By using the result in Figure \ref{spin_dt}b, we estimate the spin values, $a_{\rm est}$, for the case of various parameters of gas cloud properties, inclination angle, and dynamics (Figure \ref{spin_spin}),
where the physical meanings and fiducial values of the model parameters are shown in Table \ref{table_parameters}.

The following is a summary of what we can see from Figure \ref{spin_spin}a.
We show in this Figure the relation between $a_{\rm true}$ and $a_{\rm est}$ with the various shapes of the gas cloud, emissivity profiles, and inclination angles.
The estimated spin value, $a_{\rm est}$, is similar to that of the true value, $a_{\rm true}$, among the different thicknesses of the cross section of the gas cloud, $R_{\rm cloud}$, where the error of the spin parameter is $|\delta a/M|=|a_{\rm est}-a_{\rm true}|/M\leq 0.07$ (see red circle in Figure \ref{spin_spin}a). 
If we change the power-law index of the emissivity profile from $\Gamma=-2$ to $\Gamma=0$ and fix the other parameters, the estimated spin value is similar to the true one, where the error of the spin parameter is $|\delta a/M|\leq 0.06$ (see blue circle in Figure \ref{spin_spin}a).
The estimated spin value is not significantly altered by the change of the arc chord angle of the gas cloud, $\delta\phi$, since the time scale is determined by the period of photon rotation around the black hole (see green circle in Figure \ref{spin_spin}a).
The error of the spin is $|\delta a/M|\leq 0.2~(0.1)$ for $a/M<0.6$ ($a\geq 0.6$).
We change the inclination angle from $i=75^\circ$ to $i=40^\circ$ and fix the other parameters (see purple circle in Figure \ref{spin_spin}a).
The peak intervals between the primary and secondary peaks are not significantly altered by the change in the inclination angle, where the error of the estimated spin is $|\delta a/M|\leq 0.15$.

Next, we show in Figure \ref{spin_spin}b the $a_{\rm est}$ dependence on the motion of the gas cloud.
The larger is the initial infalling velocity of the gas cloud, the rotational beaming effect is smaller due to the radial motion. 
The radial position of the radiation region of the secondary component decreases when $|u_{\rm o}{}^1|$ increases, and the period of photon rotation decreases.
Therefore, the estimated spin values are larger than the corresponding true ones (see red circle in Figure \ref{spin_spin}b).
If the gas cloud has no angular momentum ($L/L_{\rm ms}=0$), the estimated spin value is larger than $a_{\rm true}$, since the beaming effect outside $r_{\rm ph}$ decreases and the light bending effect near the photon circular orbit is dominant (see blue circle in Figure \ref{spin_spin}b).
If the gas cloud has a high angular momentum (say, $L/L_{\rm ms}=0.95$) the peak interval increases, since the gas cloud approaches the photon circular orbit more slowly. 
Therefore, the time interval of the peaks increases with the increase of $L/L_{\rm ms}$, and so $a_{\rm est}$ is smaller than $a_{\rm true}$ (see green circle in Figure \ref{spin_spin}b).

Finally we summarize the property of the secondary peak and relation between $a_{\rm true}$ and the error, $\delta a=a_{\rm est}-a_{\rm real}$, for each true spin value (Figure \ref{spin_spin}c).
In the case of the low spin (say, $a_{\rm true}/M\leq 0.4$), the amplitude of the secondary peak is low (see Figure \ref{spin_dt}a) and we roughly estimate the spin values from the interval of the peaks for the various motions of the gas cloud.
With an increase in spin ($a_{\rm true}/M=0.6$), the frame-dragging effect enforces the amplitude of the secondary peak and aligns the values of the peak interval. The error of the estimated spin decreases ($|\delta a/M|\leq 0.20$) due to the spin alignment.
In the case of high spin values ($a_{\rm true}/M\geq 0.8$), we obtain high accuracy spin values $|\delta a/M|\leq 0.10$.

\subsection{Applications to the light curves measured with synthetic EHT observations}\label{subsec_implications_VLBI}
\begin{figure}\begin{center}\includegraphics[width=9.2cm]{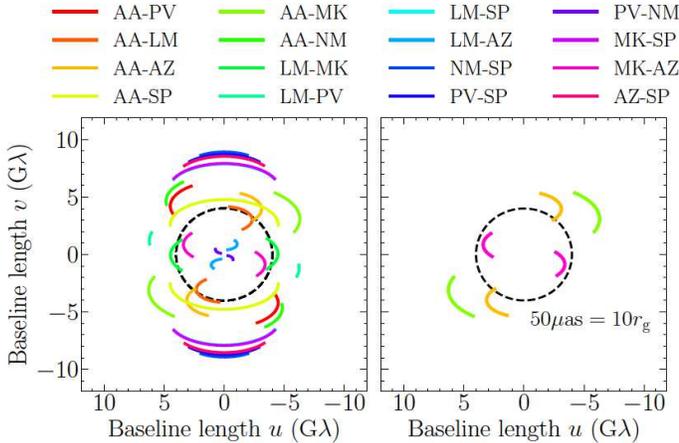} \end{center}
\caption{
Possible $(u,v)$ tracks for 16 baselines expected near-future EHT observations for Sgr\,A*; 
(a) the $(u, v)$ coverage of all baselines, where the colors of the lines of each baseline are summarized in the upper column, 
(b) a part of $(u,v)$ coverage which enables us to calibrate visibility amplitude by utilizing array redundancy of co-located telescopes (i.e., network calibration).
Axes of $(u,v)$ plane are in units of gigawavelengths at the spatial frequency, $\nu_{\rm EHT} = 230~{\rm GHz}$.
Abbreviations for each array: AA, ALMA/APEX; LM, LMT; NM, NOEMA; PV, PV 30~m; MK, JCMT/SMA; 
AZ, SMT/KP;  SP, SPT.
Here, NOEMA and KP will join the EHT Array after 2020.
}\label{uvcov}
\end{figure}

Sgr\,A* would be the most promising target for application of our approaches, since it shows rapid flux variations attributed to horizon-scale structures that can be timely and spatially resolved with near future EHT observations.
In VLBI observations, spatial and temporal fluctuations of the source intensity are imprinted in the visibility $V (t,u,v)$, which is Fourier coefficients of the source intensity, defined by
\begin{eqnarray}\label{visibility}
V (t,u,v) &=&\int \int I(t,x,y) \exp[2\pi i (ux+vy)]dxdy,
\end{eqnarray}
where $I(t,x,y)$ is the source intensity, $(x, y)$ is sky coordinates, 
$(u,v)$ is the baseline vector (so-called spatial frequency) between two antennas normalized by the observing wavelength projected to the plane perpendicular to the source direction (see \citealp{Thompson2017} for details).
The amplitude of the visibility $|V(t,u,v)|$ is so-called the correlated flux density.
Here, we adopt time-variable emission attributed to plenty of gas clouds intermittently falling onto the black hole with arbitrary initial model parameters. 
Then, we extract the spin dependence between direct and secondary components in the correlated flux densities from synthetic EHT observations, as follow in details.

\begin{enumerate}
\item 
We calculate a simulated movie of in-falliing gas clouds with random initial velocities of radial and rotational components, start times for infalling, and initial rotational angles. 
We adopt the total observational time $t_{\rm obs}=2000\,r_{\rm g}/c\sim 11~{\rm hrs}$, which is close to that of EHT observations, and the time resolution of $1r_{\rm g}/c \sim 20~{\rm s}$  larger than the typical coherence time of EHT observations ($\sim 10\ {\rm s}$; e.g.,  \citealp{Doeleman2009}). 
From previous results of velocity distributions of accretion flow based on RIAF models and GRMHD simulations (e.g., \citealp{Manmoto1997, Yuan2009, Narayan2012}), we assume that initial velocities of gas clouds has uniform random distribution within the range of $-0.2\leq u_{0}{}^r\leq0$ and $0.2\leq L/L_{\rm ms}\leq 0.95$.
We neglect the collision between each infalling gas cloud and fix the other model parameters, $R_{\rm cloud}=0.2r_{\rm g}, \delta\phi=2\pi/3$, and $i=75^\circ$, since the estimated spin values do not critically depend on them (see Figure \ref{spin_spin}a).

\item 
We calculate $V(u,v,t)$ using Equation (\ref{visibility}) and an array expecting after 2020 consisting of stations at seven sites: ALMA, APEX, LMT, NOEMA,  PV 30~m, JCMT, SMA, SMT, KP 12~m, and SPT (see \citealp{EHT2019b}, Table \ref{table_telescope} for the abbreviation of each telescope). 
In Figure \ref{uvcov} (a), we show the baseline coverages of the entire array for Sgr\,A* observations, where the $(u,v)$ coverage is calculated by the \texttt{eht-imaging} libirary (\citealp{Chael2016}; \citealp{Chael2018a}).
We especially focus on specific baselines, between Chile (ALMA/APEX), Hawaii (SMA/JCMT) and Arizona (SMT/KP 12~m) (Figure \ref{uvcov}b). 
These baselines are between sites with redundant stations, where high-accuracy calibrations of correlated-flux densities are available with network-calibrations (\citealp{Fish2011, Johnson2015, EHT2019c}).

\item 
We add thermal noises to synthetic data sets.
The noise obeys a Gaussian distribution in the complex plane with the standard deviation (e.g., \citealp{Thompson2017}), 
\begin{eqnarray}\label{noise_visibility}
\sigma_{ij} = \frac{1}{\eta} \sqrt{\frac{{{\rm SEFD}_i}{{\rm SEFD}_j}}{2\Delta \nu \Delta t}}, 
\end{eqnarray}
\noindent
where $\eta$ is the quantization efficiency (=0.88 for 2-bit sampling; \citealp{Thompson2017}), 
$\Delta \nu\,(\approx 2~{\rm GHz})$ corresponds to the bandwidth of a single band and single polarization spectral window (\citealp{EHT2019b}), $\Delta t(=20\ {\rm s})$ is the integration time, SEFD is the system equivalent flux density equal to all thermal noises produced from the telescope receiver chains, Earth's atmosphere, and astronomical background (see Table  \ref{table_telescope}), and subscripts $(i,j)$ indicate telescopes involved in the corresponding baseline.

\item 
We extract direct and secondary components from correlated flux densities composed of radiation of gas clouds and thermal noises.
In this paper, we adopt a following simple method called "superposed shot analysis" which has been developed to detect short timescale features of X-ray variation observed in black hole binaries (e.g,  \citealp{Oda1971, Negoro1994, Gierlinski2003, Focke2005, Yamada2013}). 
We divide the light curve into segments with timescale, $t_{\rm seg}$ larger than time interval between direct and secondary components ($\gtrsim 50r_{\rm g}/c$, see figure \ref{spin_dt}b), and superpose the light curves in each segment by aligning their maximum peaks and obtain the superposed light curve, $f_{\rm tot}(t)$.
We presume that most of maximum peaks in each segment correspond to direct or secondary components of infalling gas clouds, and inspect the secondary peak using following criteria: 
the difference between amplitude of the secondary component, $f_{\rm secondary}$, and the flux of the valley at the time between the maximum peak and secondary one, $f_{\rm val}$, should be more than 0.1 times as large as ${\rm max}[f_{\rm tot}(t)]-f_{\rm val}$.
If the secondary peak satisfies upper criteria, we calculate time interval between direct and secondary components, using Equation (\ref{eq_dt}).

\item 
By adopting the relation between $\delta t$ and $a/M$ (Figure \ref{spin_dt}b) to the time interval estimated from the simulation result, we estimate the spin value, $a_{\rm est}$. Procedures 1-5 is the single set of the synthetic observation.
\item 
We repeat the synthetic observation for 1000 times, estimate the distribution of $a_{\rm est}$, and then evaluate its $1\sigma$ confidence range.
We count the number of detections of secondary peaks based on the Procedure 4 during the iteration, and call it detectability. 
\end{enumerate}

\begin{deluxetable*}{llr}
\tablenum{2}
\tablecaption{Stations in simulated observations \label{table_telescope}}
\tablewidth{0pt}
\tablehead{
\colhead{Telescope \ \ \ \ \ \ \ \ \ \ \ \ \ \ \ \ \ \ \ \ \ \ \ \ \ } & \colhead{Abbreviations for each array  \ \ \ \ \ \ \ \ \ \ \ \ \ \ \ \ \ \ \  } & \colhead{SEFD (Jy)}
}
\startdata
ALMA	& AA	& 94\\
LMT		& LM     & 570\\
NOEMA & NM & 1500 \\
PV 30~m & PV & 1400 \\
JCMT/SMA	& MK	& 4700\\
SMT	& AZ & 11000\\
SPT	& SP	& 9000\\
\enddata
\end{deluxetable*}


To perform synthetic EHT observations for Sgr\,A*, we need two additional physical parameters: a number of infalling gas clouds per unit time, $\dot{N}$, and synchrotron emissivity, $j_{\rm synch}$. 
We estimate them by setting values of the gas density, electron temperature and magnetic field are $\rho_{\rm gas}\sim 10^{-16}~{\rm g\ cm^{-3}}$, $T_{\rm e}\sim 10^{10}~{\rm K}$, and $B\sim 100~{\rm G}$, respectively, often seen in RIAF models or GRMHD simulations in the past literature (\citealp{Yuan2009, Goldstone2005, Moscibrodzka2009, Dexter2010, Narayan2012, Chael2018b}).
If the mass accretion is mainly due to the infalling gas clouds, the number of the infalling gas cloud per unit time is expressed as $\dot{N}\sim \dot{M}/(\rho_{\rm gas}V_{\rm cloud})\sim 0.2 c/r_{\rm g}$, 
where the mass accretion rate is $\dot{M}\sim10^{-8}~M_{\odot}{\rm yr}^{-1}$ (\citealp{Agol2000, Bower2000, Marrone2007, Bower2018}),
$V_{\rm cloud}(\approx  g_{rr}\pi\delta\phi r_{\rm i}R_{\rm cloud}^2\sim 10^{36}\ {\rm cm^{3}})$ is the volume of each gas cloud, and $g_{rr}$ is the radial metric component in Boyer-Lindquist coordinates.
The synchrotron emissivity, $j_{\rm synch}(\nu_{\rm EHT})\sim 4\times 10^{-15}\ {\rm erg\ cm^{-3}s^{-1}\ str^{-1}/ Hz^{-1}}$, is calculated with Equation (\ref{eq_synchrotron_emissivity}) and typical parameters of RIAF, and GRMHD simulation results.
Note that synchrotron self-absorption of the gas clouds can be neglected, since the optical depth of the synchrotron radiation, $\tau_{\rm synch}$, is sufficiently small; $\tau_{\rm synch}\approx j(\nu_{\rm EHT})R_{\rm cloud}/[B_{\nu_{\rm EHT}}(T)]\sim 0.05 \ll 1$ due to the small size and low emissivity of the infalling gas cloud, 
where $B_{\nu}(T)$ is the Planck function (see \citealp{Manmoto1997}).
\begin{figure*}\begin{center}\includegraphics[width=18 cm]{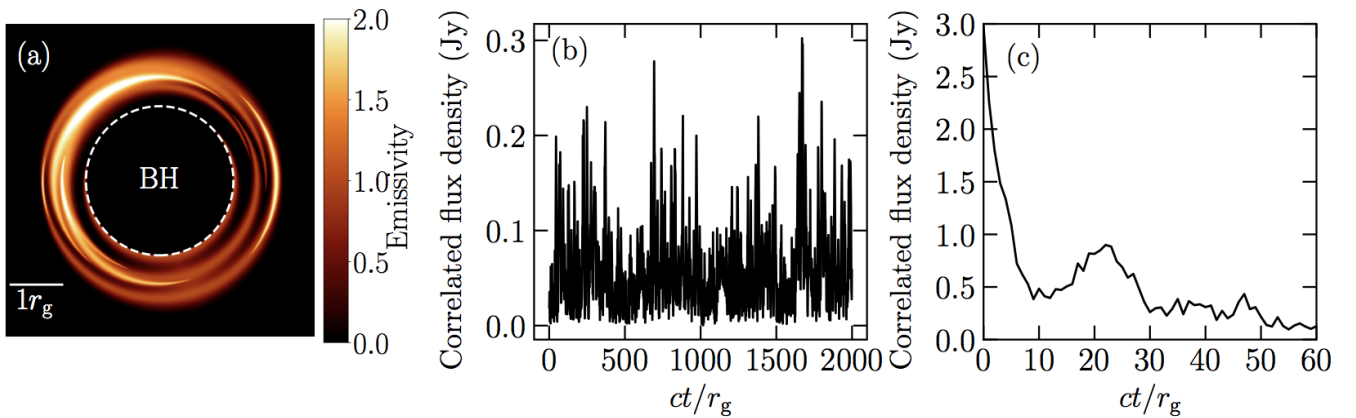} \end{center}
\caption{
Synthetic EHT observation for infalling gas clouds for the case with $(a/M,i)=(0.9,75^{\circ})$: (a) spatial emissivity distribution of the infalling gas clouds, where the origin is set to be at the
center of the black hole, the white dotted circle indicates the event horizon, and the emissivity is normalized by synchrotron emissvity, ($j_{\rm synch}\sim 4\times 10^{-15}\ {\rm erg\ cm^{-3}s^{-1}\ str^{-1}/ Hz^{-1}}$.
(b) simulated light curve of the correlated flux density on the ALMA-JCMT/SMA baseline, where high accuracy measurements are available using the network calibration, and 
(c) the example of the superposed light curve by adopting the superposed shot analysis to the light curve shown in panel (b), where the time duration of each segment is $t_{\rm seg}(=100r_{\rm g}/c)$.
}\label{syntheticobs_lightcurve}
\end{figure*}


\begin{figure}\begin{center}
\includegraphics[width=9.1cm]{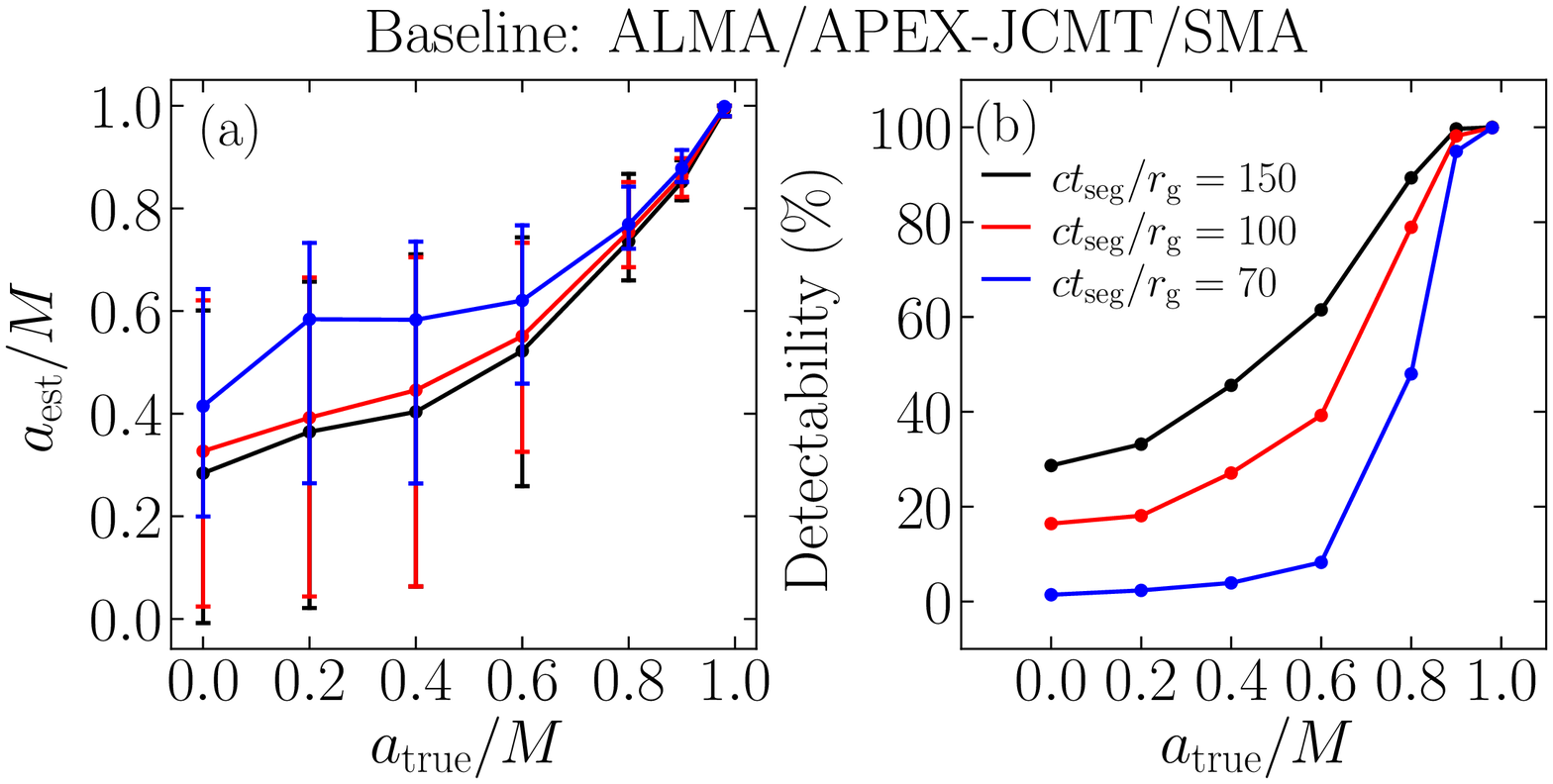} \end{center}
\caption{
The accuracy of the estimated black hole spin and detectability of the secondary peak.
Panel (a) shows the estimated spin values as functions of real spin values, $a_{\rm true}$, by adapting the relation between $\delta t$ and $a/M$ (Figure \ref{spin_dt}b) to the simulaetd light curve on the ALMA--JCMT/SMA baseline (Figure \ref{syntheticobs_lightcurve}).
Here, each curve corresponds to time segments of $ct_{\rm seg}/r_{\rm g}=150$ (black), $100$ (red), and $70$ (blue), respectively, and vertical bars of each point indicate its 1$\sigma$ confidence ranges of the estimated spin values.
Panel (b) plot the detectability of the secondary peak by using the criterion of the synthetic observation (see Procedure 4 in Section \ref{sec_discussion}).
}\label{syntheticobs}
\end{figure}

%
We show in Figure \ref{syntheticobs_lightcurve} a single set of a synthetic observational result.
Figure \ref{syntheticobs_lightcurve}a shows a spatial distribution of the emissivity of the infalling gas clouds on the equatorial plane at a certain time during the synthetic observation, where the horizontal and vertical axes correspond to the Boyer Lindquist coordinates.
We randomly choose the start times of the infall of $\dot{N}\ [r_{\rm g}/c]\times 2000\ [c/r_{\rm g}]=400$ gas clouds during the time interval between  $0$ and $2000 r_{\rm g}/c$ and the number of gas clouds in Figure \ref{syntheticobs_lightcurve}a is $\dot{N}t_{\rm infall}\sim 4$.
The emissivity is normalized by $j_{\rm synch}(\nu_{\rm EHT})$.
Plenty of gas clouds intermittently fall onto the event horizon denoted by the white dotted circle and originate complex flux variation.

We show in Figure \ref{syntheticobs_lightcurve}b a simulated light curve for the case with $(a/M, i)=(0.9, 75^\circ)$.
An interesting baseline is between Chile (ALMA/APEX) and Hawaii (JCMT/SMA), since it provides us accurately calibrated correlated-flux densities due to network-calibrations and have been used in EHT observations of Sgr\,A* since 2017.
The light curve of the simulated correlated flux density on the ALMA-JCMT/SMA baseline, composed of peaks attributed to gas clouds with various initial velocities, start times for infalling, and initial rotational angles. 
The peak flux of the light curve is $\sim 0.05-0.3\ {\rm Jy}$, which is the same order of radio flux variations of Sgr\,A* in the quiescent state (\citealp{Yusef-Zadeh2009}), and so the evaluation of the synchrotron emission of infalling gas clouds is reasonable for the real observational results of Sgr\,A*.
We adopt the superposed shot analysis explained in Procedure 4; we divide the light curve shown in Figure \ref{spin_dt}b into segments with timescale larger than the time interval between direct and secondary components ($ct_{\rm seg}/r_{\rm g}=100$), and superpose the light curves in each segment by aligning their maximum peaks and obtain the superposed light curve (see Figure \ref{spin_dt}c).
We found that the systematic features of the direct and secondary components are detected and the time interval between each peak is $c\delta t/r_{\rm g}=20$, which is consistent with the result of Section \ref{sec_results}.

In Figure \ref{syntheticobs}a, we show the spin values as functions of real spin values, $a_{\rm true}$, by adapting the relation between $\delta t$ and $a/M$ (Figure \ref{spin_dt}b) to the superposed light curves (Figure \ref{syntheticobs_lightcurve}b), where 
each curve corresponds to time segments of $ct_{\rm seg}/r_{\rm g}=150$ (black), $100$ (red), and $70$ (blue), respectively, 
and vertical bars of each point indicate its 1$\sigma$ confidence ranges of the estimated spin values.
In the case of the low spin (say, $a_{\rm true}/M= 0.0- 0.6$), we roughly estimate the spin values from the time interval of the peaks ($\delta a/M< 0.6$). 
If the spin increases ($a_{\rm true}/M\geq 0.8$), the frame-dragging effect aligns the values of the peak interval, and so the error of the estimated spin decreases ($\delta  a/M< 0.1$).
The relation between the true and estimated spin values does not significantly depend on the selection of the time segment.

The secondary peak is not always detected during the iteration processes and the detectability of secondary peaks depends on the spin value.
We show in Figure \ref{syntheticobs}b the detectability based on the criterion in Procedure 4 in this section.
Let us focus on the case with time segment of $ct_{\rm seg}/r_{\rm g}=100$ (red curve) to investigate the relation between the true spin and detectability.
In the case of the high spin ($a/M=0.8-0.95$), [low spin ($a/M=0-0.6$)] the detectability is higher than $80~\%$ (lower than $40~\%$), since the frame-dragging effect strongly (hardly) enforces the amplitude of the secondary peaks (see Figure \ref{spin_dt}a).
Therefore, if the spin is $a/M\leq 0.6$, we hardly observe the secondary peak and constrain the low spin value using the low detectability and variation of estimated spin values. 
For the relatively high spin values ($a/M> 0.6$), we can detect a secondary peak and estimate spin values from the time interval of each peak.
The tendency between the true spin and detectability does not depend on the selection of the time segment.
We summarize in Table \ref{table_spin_baseline} the accuracy of the estimated spin values for the case with each baseline. 
The accuracy of the estimated spin values are not significantly different each other, and so we can estimate the spin values with the similar accuracy by using each baseline.

\begin{deluxetable*}{cccc}
\tablenum{3}
\tablecaption{Estimated spin value of each baseline\label{table_spin_baseline}}
\tablewidth{0pt}
\tablehead{
\colhead{baselines \ \ \ \ \ } & 
\colhead{$a_{\rm est}/M$ for $a_{\rm true}/M=0.2$} &
\colhead{$a_{\rm true}/M=0.6$} & 
\colhead{$a_{\rm true}/M=0.9$}
}
\decimalcolnumbers
\startdata
ALMA-JCMT/SMA	& 0.39 (0.10-0.66) & 0.56 (0.33-0.77) & 0.86 (0.82-0.90) \\ 
ALMA-SMT			& 0.33 (-0.00-0.62) & 0.56 (0.31-0.76) & 0.87 (0.82-0.90) \\ 
JCMT/SMA-SMT		& 0.34 (0.04-0.62) & 0.62 (0.37-0.80) & 0.89 (0.83-0.94) \\ 
\hline
ALMA- PV 30~m	& 0.27 (-0.03-0.59) & 0.57 (0.34-0.75) & 0.86 (0.82-0.90) \\ 
ALMA-LMT			& 0.30 (0.01-0.61) & 0.55 (0.31-0.75) & 0.86 (0.82-0.90) \\ 
ALMA-SPT			& 0.31 (0.01-0.61) & 0.54 (0.28-0.73) & 0.86 (0.82-0.90) \\ 
ALMA-NOEMA		& 0.31 (-0.01-0.62) & 0.56 (0.34-0.76) & 0.87 (0.82-0.90) \\ 
LMT-JCMT/SMA		& 0.33 (0.02-0.63) & 0.56 (0.34-0.74) & 0.87 (0.84-0.92) \\ 
LMT- PV 30~m		& 0.29 (-0.01-0.62) & 0.56 (0.34-0.75) & 0.86 (0.82-0.90) \\ 
LMT-SPT		    		& 0.33 (0.04-0.62) & 0.57 (0.33-0.77) & 0.87 (0.84-0.92) \\ 
LMT-SMT				& 0.31 (0.02-0.60) & 0.57 (0.31-0.75) & 0.87 (0.84-0.92) \\ 
NOEMA-SPT			& 0.35 (0.03-0.63) & 0.58 (0.32-0.77) & 0.88 (0.84-0.92) \\ 
PV 30~m-SPT 		& 0.35 (0.05-0.67) & 0.58 (0.34-0.77) & 0.88 (0.84-0.92) \\ 
PV 30~m-NOEMA	&0.31 (0.02-0.60) & 0.56 (0.31-0.75) & 0.87 (0.84-0.92) \\ 
JCMT/SMA-SPT		& 0.38 (0.05-0.68) & 0.62 (0.36-0.81) & 0.89 (0.84-0.94) \\ 
SMT-SPT			 	& 0.39 (0.05-0.69) & 0.65 (0.40-0.84) & 0.91 (0.85-0.97) \\ 
\enddata
\tablecomments{
Top three lines show the estimated spin values for redundant site, and bottom lines corresponds to other ones.
}
\end{deluxetable*}


\section{Discussion}\label{sec_discussion}

\begin{figure}[ht!]
\plotone{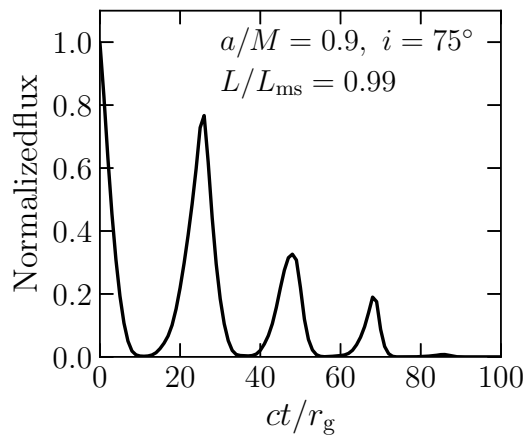}
\caption{
The superposed light curve in the case of a high angular momentum, $L/L_{\rm ms}=0.99$.
}\label{high_L}
\end{figure}

\subsection{Uncertainty in high angular momentum cases}\label{subsec_highL}
In the case of the high angular momentum of the gas cloud ($L/L_{\rm ms}\approx 1$), it may rotate around the black hole at least twice.
For the extreme angular momentum, $L/L_{\rm ms}=0.99$, the superposed light curve has two peaks due to the direct component (Figure \ref{high_L}).
The time and azimuthal angle of the photon rotation at each peak are 
$(ct/r_{\rm g}, \langle \phi_{\rm ray}/2\pi\rangle)=(0, 0.46), (26, 0.53), (47, 1.10), (68, 1.39)$, and $(86, 2.56)$, 
respectively, and so the first two peaks are generated by the direct radiation.
The time intervals of each peak are $c\delta t/r_{\rm g}=26, 21, 21$, and $18$, respectively.
The contribution of the direct component may obscure the estimation result of the spin value.
Though we may not consider the gas cloud with the high angular momentum, since the RIAF models predict the accretion disk with the sub-Keplerian orbital velocity, it may be necessary to construct the method to distinguish the direct component and  secondary one for general cases.

\begin{figure*}\begin{center}
\includegraphics[width=17cm]{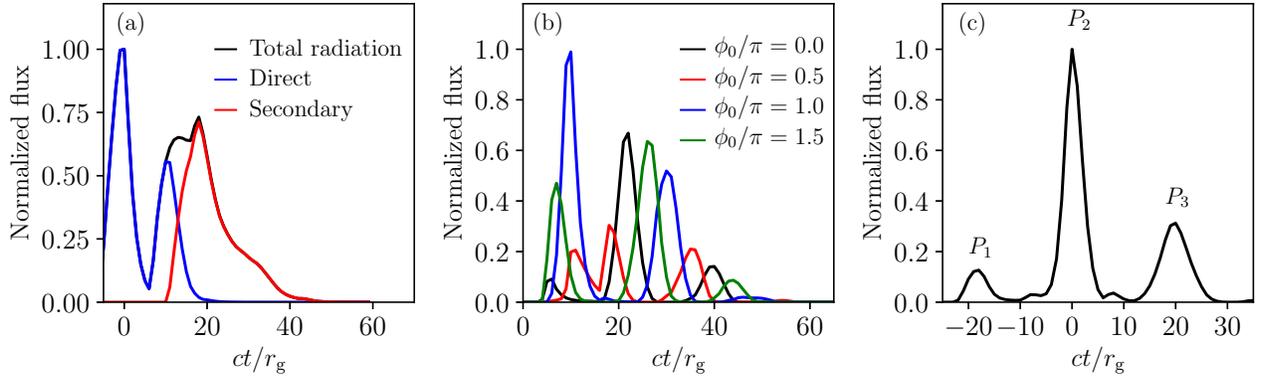} \end{center}
\caption{
Flux variation of infalling gas clouds where the initial azimuthal angle is $\theta_{0}=60^{\circ}$, inital angular velocity is $\Omega = 0.2 \Omega_{\rm ms}$, and other parameters are the same as those in Figure \ref{ring}: 
(a) time variation of the normalized flux, where black, blue, and red curves are for total, direct, and secondary components, respectively, 
(b) light curves of the arc-shaped gas cloud with the various initial azimuthal angle,$\phi_{0}=0,0.5, 1.0, 1.5$, where the flux is normalized by the maximum one in the case of $\phi_{0}=1.0$, 
(c) superposed light curve arc-shaped gas cloud with 20 initial azimuthal angles, $\phi_{0}/(2\pi)=[0,19/20]$, by aligning their maximum peaks for the case of each spin.
}\label{polar}
\end{figure*}

\subsection{Dependence on initial polar angles}
We constructed a method for black hole spin measurement under a situation that a gas cloud is on the equatorial plane and does not have an  polar angular velocity.
However, alignment between the accretion disk and black hole spin axis is not necessary (e.g., \citealp{Dexter2013}).
In such a situation, gas clouds may fall toward the black hole with motions to the polar direction and observational radiation is affected by its relativistic effects. 
In this subsection, we investigate additional properties of flux variation with an motion to the polar direction.

We show in Figure \ref{polar}a an example of flux variation of infalling gas clouds, where the initial polar angle from the axis of the black hole spin is $\theta_{0}=60^{\circ}$, initial polar angular velocity is zero, initial angular velocity is $\Omega = 0.2 \Omega_{\rm ms}$, and other parameters are the same as those in Figure \ref{arcgas}.
The total flux (the black curve) is composed of the direct (blue curve) and secondary (red curve) components and the light curve has three characteristic peaks.
During $0\leq ct/r_{\rm g}\leq 13$, the direct component is dominant. At the time of $ct/r_{\rm g}=0$, the radiation comes from $[\langle r_{\rm ray}/r_{\rm g}\rangle, \langle \phi_{\rm ray}/(2\pi)\rangle, \langle \theta_{\rm ray}/\pi\rangle] = (2.08, 0.41, 0.34)$. It indicates that the component is constructed by the rotational beaming effect of the direct component.
The first component rapidly decreases due to the combination of gravitational redshift, capture of photons into the black hole, and  Doppler deboosting to the polar direction.
At the time of $ct/r_{\rm g}=11$, there is an additional peak component comes from $[\langle r_{\rm ray}/r_{\rm g}\rangle, \langle \phi_{\rm ray}/(2\pi)\rangle, \langle \theta_{\rm ray}/\pi\rangle] = (2.09, 0.83, 0.34)$, and so it is first radiation without strong polar deboosting effect.
After $ct/r_{\rm g}>13$, secondary component is dominant; at the time of $ct/r_{\rm g}=19$, the radiation peak whose main component comes from $[\langle r_{\rm ray}/r_{\rm g}\rangle,\langle \phi_{\rm ray}/(2\pi)\rangle, \langle \theta_{\rm ray}/\pi\rangle] = (2.03, 1.27, 0.34)$.

The important feature for the case is that the magnitudes of direct and secondary peaks are comparable each other, since the direct component is affected by the polar de-boosting effect.
In the case of arc-shaped gas clouds, the direct peak is sometimes smaller than that of the secondary component (see the green curve in Figure \ref{polar}b).
If we align the maximum peaks for various case of $\phi_{0}$, such components make an additional peak, P1, in Figure \ref{polar}c.
The time interval between $P_1$ and $P_2$ is $c\delta t/r_{\rm g}=18$, which is similar to that between $P_2$ and $P_3$ ($c\delta t/r_{\rm g}=20$).
This additional feature may inform us the information for motion {to the polar direction} and structure of infalling gas clouds.
The more detail investigation requires consideration of time fluctuation of  polar structure of gas clouds and their motion, and so we investigate the effect in the future study based on GRMHD simulations.

\begin{figure*}\begin{center}\includegraphics[width=18cm]{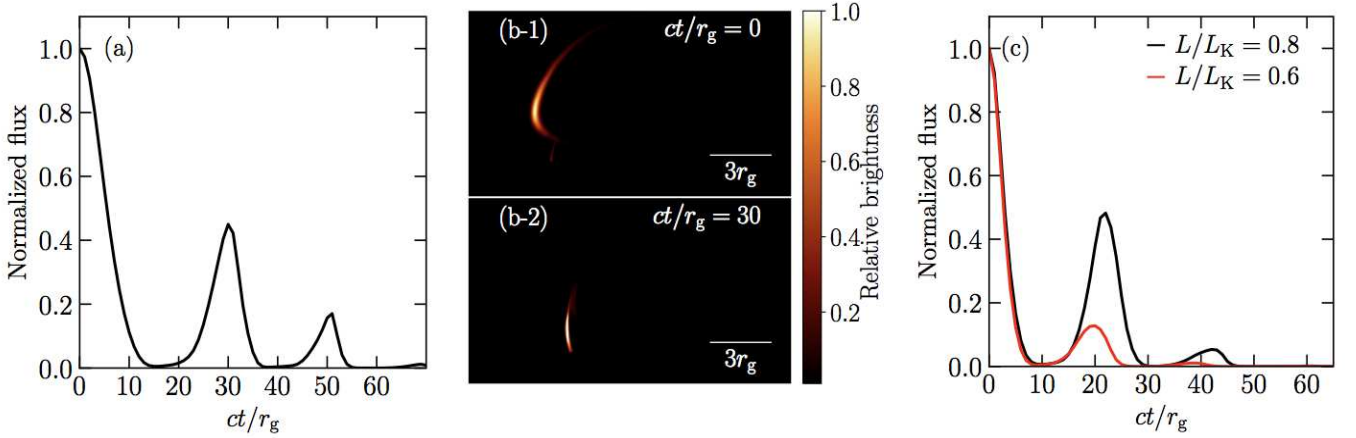} \end{center}
\caption{
Flux variation of the gas cloud falling from the accretion disk whose radius of the inner edge is  $r_{\rm i}=6r_{\rm g}$, 
where we assign the same parameters as Figure \ref{arcgas}a except the initial azimuthal angle and angular momentum. 
Panel (a) shows that the light curve of the infalling gas cloud with $\phi_{\rm o}=\pi/2$ and $L/L_{\rm K}=0.8$, where $L_{\rm K}$ is the angular momentum of the Kepler rotation at the initial radius of the infalling gas cloud ($\approx 6r_{\rm g}$).
Panel (b) is the snapshot images of the first ($ct/r_{\rm g}=0$) and second peaks ($ct/r_{\rm g}=32$).
The brightest points of each image are $(x_{\rm obs},y_{\rm obs},r_{\rm obs})=(-5.1, 1, 5.1)$ and $(-3.1, -0.1, 3.1)$, respectively.
We note that the discontinuity point of the first image $[(x_{\rm obs},y_{\rm obs})=(-3.9,-1)]$ is the component that reaches the observers through the $z<0$ region.
Panel (c) depicts the superposed light curve by using the radiation within the range of $r_{\rm obs}\leq 4$, 
where $r_{\rm obs}=\sqrt{x_{\rm obs}^2+y_{\rm obs}^2}$, and we assign $L/L_{\rm K}=0.8, 0.6$.
}\label{inner_disk}
\end{figure*}

\subsection{Uncertainty in the inner edge of the accretion disk}\label{subsec_inner_rad}
In the previous section, we assumed that the inner edge of the disk is at the radius of the marginally stable orbit, $r_{\rm ms}$, but the assumption remains a controversial issue (see \citealp{Shidatsu2014}).
Now we consider in Figure \ref{inner_disk} the situation that the inner edge of the accretion disk is $r_i/r_{\rm g}=6$ for the case of $a/M=0.9, i=75^\circ, R_{\rm cloud}/r_{\rm g}=0.2, \Gamma=-1, L/L_{\rm K}=0.8$, where $L_{\rm K}$ is the angular momentum of the Kepler rotation at the initial radial position of the infalling gas cloud ($=0.98\,r_{\rm i}$).
At first, we investigate light variation of the arc-shaped gas cloud falling from an initial azimuthal angle, $\phi_{\rm o}=\pi/2$ (Figure \ref{inner_disk}a).
The time interval between the first and second peaks is $32 cr_{\rm g}$, which is larger than the period of photon rotation ($cT/r_{\rm g}= 20$), since the radiation position of the first peak is larger than $r_{\rm ms}$. 
The large time interval prevents us from extracting the rotation period of the photon and estimating the accurate spin value.

The contribution of the outer region may be removed by using images of VLBI observations.
We focus on the radial position of the maximum brightness on the observer's plane: 
the maximum radiation of the first peak reaches  $r_{\rm obs}\left(=\sqrt{x_{\rm obs}^2+y_{\rm obs}^2}\right)= 5.1r_{\rm g}$, while that of the second peak reaches at $r_{\rm obs}=3.1 r_{\rm g}$ (Figures \ref{inner_disk}b1 and b2). 
We assume that we detect the radiation within the range of $0\leq r_{\rm obs}/r_{\rm g}\leq 4$, and superpose light curves by adapting the superposition method to flux variation within the radius $r_{\rm obs}=4\,r_{\rm g}$ (Figure \ref{inner_disk}c), where $L/L_{\rm K}=0.8$ (black curve), or $0.6$ (red).
The peak interval of the first and second peaks is $c\delta t/r_{\rm g} =22\ (20)$ for $L/L_{\rm K}=0.8\ (0.6)$.
Using Figure \ref{spin_dt}b, we estimate the spin value from the light curve: $a/M=0.85\ (0.9)$ and the error is $\delta a/M=0.05\ (0.00)$ for the case with $L/L_{\rm K}=0.8\ (0.6)$.
Therefore, we may adapt our methodology to the case of $r_{\rm i}>r_{\rm ms}$.

The extraction will be performed in the following way based on imaging techniques.
First, we reconstruct the averaged image of the accretion disk using time-averaged and/or dynamical imaging techniques (\citealp{Lu2016, Johnson2017, Bouman2017}).
Then, we subtract the visibility component of the accretion disk from the observation including all components, and obtain the radiation of infalling gas clouds from reconstructed movies with dynamical imaging.
We may detect the flux variation of the infalling gas cloud directly.

In this paper, we focus on the time variation of infalling gas clouds and do not consider the situation that the intrinsic variability emission or lumpiness in the inner accretion disk dominates the radiation of infalling gas clouds. 
The limitation may be solved if we extract the infalling gas cloud components from the whole variation using this image domain analysis. 
In order to test the applicability of our method to such a case, it is necessary to perform synthetic observations and imaging based GRMHD simulation. 
This is our next scope of a future paper.

\section{Summary and future study}\label{sec_summary}
In this paper, we have proposed a new method for spin measurement based on spatially-resolved temporal variation of infalling gas clouds, which will be obtained by near-future VLBI observations.
We find that the light curve of the infalling gas cloud has two characteristic peaks due to the direct and  secondary radiation, and the time interval between the two peaks is determined by the period of photon rotation around the photon circular orbit.
\begin{figure}
\plotone{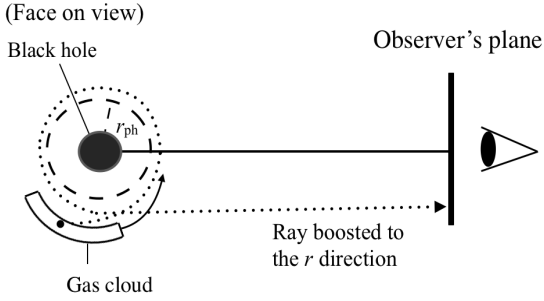}
\caption{
Schematic picture of the secondary radiation that is reflected around the photon circular orbit and then reaches the observer plane.
At the emission point, the fraction of photons is Doppler boosted to the radial direction, since the radial direction of the trajectories is the same as that of the infalling gas cloud. 
}\label{rboost}
\end{figure}
We emphasize that the radial Doppler effect of the gas cloud contributes to the enhancing the peak of the secondary component.
The fraction of the observed photon constructing the secondary peak propagates toward the black hole, and is reflected just outside the photon circular orbit due to the gravitational bending effect (Figure \ref{rboost}).
The radiation component is Doppler boosted radially, since the radial directions of the trajectories of the photons are the same as that of the gas cloud.
Therefore, the amplitude of the secondary component increases due to the radial beaming effect.

We have applied our method of spin measurement to synthetic EHT observations of Sgr\,A*.
If the black hole has $a\leq 0.6$, we can constrain the low spin value using low detectability of the secondary components and variation of estimated spin values.
In the case of high spin ($a/M>0.6$), we can detect a secondary peak and accurately estimate spin values from the time interval of each peak, since it does not depend on the shape and angular momentum of the gas cloud, radiation profile, and inclination angle.
If the radius of the inner edge of the accretion disk is larger than that of the marginally stable orbit, we may be able to estimate the black hole spin values by identifying the inner region $(r_{\rm obs}/r_{\rm g}\leq 4)$ and detect the contribution of the gas cloud in the vicinity of the black hole.

Finally, we summarize future issues.
To investigate the realistic behavior of infalling gas clouds around the supermassive black hole, we need to  examine radiative and dynamical properties of the gas clouds with 3D-GRMHD simulations. 
It is also important to consider the self-absorption effect to the secondary radiation based on realistic infalling gas clouds properties.
It is also necessary to apply the dynamical imaging and averaged imaging methods to synthetic observational results of the 3D-GRMHD simulation, and establish a method for extracting the flux variation of infalling gas clouds from total flux including the component of the accretion disk.
We evaluate the accuracy of the estimated spin value by adapting our methodology to spatially-resolved flux variation of the synthetic observational data.
These future studies will enable us to obtain an independent spin measurement of  Sgr A* by applying our methodology to the EHT observations expected in 2017-2020.

We are grateful to Vincent L. Fish and Yosuke Mizuno for many constructive and meaningful comments.
This research is supported in part by JSPS Grant-in-Aid for JSPS Research Fellow (JP17J08829).
KA is a Jansky Fellow of the National Radio Astronomy Observatory. 
The National Radio Astronomy Observatory is a facility of the National Science Foundation operated under cooperative agreement by Associated Universities, Inc. 
The Black Hole Initiative at Harvard University is financially supported by a grant from the John Templeton Foundation. 
K.Akiyama is financially supported in part by a grant from the National Science Foundation (AST-1614868).

\appendix
\section{Angular velocity of rays around photon circular orbit}\label{sec_appendix}

In this paper, we consider the rays rotating around the Kerr black hole.
In Boyer-Lindquist coordinates, the black hole spacetime is
\begin{eqnarray*}
ds^2=g_{00}dt^2+2g_{03}dtd\phi +g_{11}dr^2+g_{22}d\theta^2 +g_{33}d\phi^2,
\end{eqnarray*}
where we denote $g_{\mu \nu}$ as the metric components, and shall use units in which $G=c=1$ in this appendix (\citealp{Bardeen1972}).
The trajectory of a photon is described by $ds^2 =g_{\mu\nu}dx^{\mu}dx^{\nu}=0$.
 On the equatorial plane, the angular velocity, $\Omega=d\phi/dt$, around the photon circular orbit is expressed as
\begin{eqnarray}\label{Omega_ray}
\Omega
&=& \left(r^2 +a^2+\frac{2Ma^2 }{r}\right)^{-1}\left(\frac{2Ma}{r}+\sqrt{r^2-2Mr+a^2}\right).
\end{eqnarray}
The period corresponding to the angular velocity, $T$, is written as
\begin{eqnarray}\label{T_ray}
T=  2\pi\left(r^2 +a^2+\frac{2Ma^2 }{r}\right)\left(\frac{2Ma}{r}+ \sqrt{r^2-2Mr+a^2}\right)^{-1}.
\end{eqnarray}

\end{document}